\documentclass[12pt]{article}

\usepackage{graphicx}
\usepackage{epsf,amsfonts,hyperref}
\renewcommand{\appendix}[1]{
    \addtocounter{section}{1}
    \setcounter{equation}{0}
    \renewcommand{\thesection}{\Alph{section}}
    \section*{Appendix \thesection\protect\indent #1}
    \addcontentsline{toc}{section}{Appendix \thesection\ \ \ #1}
}
\newcommand\encadremath[1]{\vbox{\hrule\hbox{\vrule\kern8pt
\vbox{\kern8pt \hbox{$\displaystyle #1$}\kern8pt}
\kern8pt\vrule}\hrule}}
\def\enca#1{\vbox{\hrule\hbox{
\vrule\kern8pt\vbox{\kern8pt \hbox{$\displaystyle #1$}
\kern8pt} \kern8pt\vrule}\hrule}}

\newcommand\figureframex[3]{
\begin{figure}[bth]
\hrule\hbox{\vrule\kern8pt
\vbox{\kern8pt \vbox{
\begin{center}
{\mbox{\epsfxsize=#1.truecm\epsfbox{#2}}}
\end{center}
\caption{#3}
}\kern8pt}
\kern8pt\vrule}\hrule
\end{figure}
}
\newcommand\figureframey[3]{
\begin{figure}[bth]
\hrule\hbox{\vrule\kern8pt
\vbox{\kern8pt \vbox{
\begin{center}
{\mbox{\epsfysize=#1.truecm\epsfbox{#2}}}
\end{center}
\caption{#3}
}\kern8pt}
\kern8pt\vrule}\hrule
\end{figure}
}

\renewcommand{\thesection}{\arabic{section}}

\makeatletter
\@addtoreset{equation}{section}
\makeatother
\newtheorem{theorem}{Theorem}[section]

\newtheorem{remark}{Remark}[section]
\newtheorem{proposition}{Proposition}[section]
\newtheorem{lemma}{Lemma}[section]
\newtheorem{corollary}{Corollary}[section]
\newtheorem{definition}{Definition}[section]
\def\br{\begin{remark}\rm\small}
\def\er{\end{remark}}
\def\bt{\begin{theorem}}
\def\et{\end{theorem}}
\def\bd{\begin{definition}}
\def\ed{\end{definition}}
\def\bp{\begin{proposition}}
\def\ep{\end{proposition}}
\def\bl{\begin{lemma}}
\def\el{\end{lemma}}
\def\bc{\begin{corollary}}
\def\ec{\end{corollary}}
\def\beaq{\begin{eqnarray}}
\def\eeaq{\end{eqnarray}}
\newcommand{\proof}[1]{{\noindent \bf proof:}\par
{#1} $\square$
\bigskip}

\newcommand{\eq}[1]{Eq.~(\ref{#1})}

\newcommand{\beq}{\begin{equation}}
\newcommand{\eeq}{\end{equation}}
\newcommand{\bea}{\begin{eqnarray}}
\newcommand{\eea}{\end{eqnarray}}

\newcommand{\vs}{\vspace{0.7cm}}
\renewcommand{\and}{{\qquad {\rm and} \qquad}}

 \newcommand{\Tr}{{\,\rm Tr}\:}

\newcommand{\Res}{\mathop{\,\rm Res\,}}

\renewcommand{\l}{\lambda}
\newcommand{\Pint}{{\int\kern -1.em -\kern-.25em}}

\renewcommand{\Re}{{\mathrm{Re}}}
\renewcommand{\Im}{{\mathrm{Im}}}

\renewcommand{\l}{\lambda}

\begin{document}
\sloppy


\pagestyle{empty}

\addtolength{\baselineskip}{0.20\baselineskip}
\begin{center}
\vspace{26pt}
{\large \bf {Gaussian matrix model in an external field and non-intersecting Brownian motions}}
\newline
\vspace{26pt}

{\sl N. \, Orantin}\hspace*{0.05cm}\footnote{ E-mail: nicolas.orantin@uclouvain.be}\\
\vspace{6pt}
D\'epartement de math\'ematiques,\\
Chemin du cyclotron, 2\\
Universit\'e Catholique de Louvain,\\
1348 Louvain-la-Neuve, Belgium.\\
\end{center}

\vspace{20pt}
\begin{center}
{\bf Abstract:}
We study the Gaussian hermitian random matrix ensemble with an external matrix which has an arbitrary number of eigenvalues with arbitrary
multiplicity. We compute the limiting eigenvalues correlations when the size of the matrix goes to infinity in non-critical
regimes. We show that they exhibit universal behavior and can be expressed with the Sine and Airy kernels. We also briefly
review the implication of these results in terms of non-intersecting Brownian motions.

\end{center}

\vspace{26pt}
\pagestyle{plain}
\setcounter{page}{1}

\tableofcontents

\section{Introduction and statement results}
In these notes, we are interested in the hermitian  matrix model in an external field defined by the partition function:
\beq
Z:= \int_{H_N} dM e^{-N \Tr \left({M^2 \over 2} - A M \right)}
\eeq
where one integrates over the hermitian matrices $M$ of size $N \times N$, $A$ is a deterministic diagonal matrix of the form
\beq
A := \hbox{diag}(\overbrace{a_1, \dots,a_1}^{n_1},\overbrace{a_2,\dots,a_2}^{n_2}, \dots ,\overbrace{a_k,\dots,a_k}^{n_k})
\eeq
with  $a_i\neq a_j$ for $i \neq j$ and the measure $dM$ is the product of the Lebesgue measure of the real entries of the matrix M:
\beq
dM := {\displaystyle \prod_{i<j}} d\Re\left(M_{ij}\right) d\Im\left(M_{ij}\right) {\displaystyle \prod_i} d M_{ii}.
\eeq
The main motivation of this work comes from the study of the exclusion process defined as follows. Consider $N$ independent
non-intersecting Brownian motions
$\{x_i(t)\}_{i=1}^N$ starting from $x=0$ at time $t=0$ and ending at point $x=a_i(1)$ at $t=1$ for $n_i$ of them
with $i=1, \dots, k$. One obviously imposes $\sum_{i=1}^k n_i = N$. Indeed, it is known \cite{TW,Bleher} that,up to a rescaling,
\beq
m_i={x_i \over t (1-t)},
\eeq
the position of the particles $x_i(t)$ at any time $t \in (0,1)$, are distributed as the eigenvalues $m_i$ of the random matrix $M$
submitted to an external matrix $A(t)$ whose eigenvalues are given by
\beq
a_i(t)=a_i(1) \sqrt{t \over 1-t}.
\eeq
We study the behavior of this process when the number $N$ of Brownian motions gets very large, i.e. the size of the random
matrix $M$ gets large and the filling fractions ${n_i \over N}=\epsilon_i$ are kept fixed\footnote{In order to impose this condition,
one has to restrict our study to $N$ common multiple of the denominators of the fractions $\epsilon_i$. It should be possible
to release this constraint by showing that the corrections one should get by adding an isolated eigenvalue to $A$ are not seen
at the leading order of the physical quantities when $N\to \infty$. This issue is not essential for the present work and is
not studied here.}. In this limit, at a given time $t$ the Brownian motions accumulate to fill a set of $l(t)$ intervals
with $1\leq l(t)\leq k$ and the behavior near the boundaries:
$l(t)\to 1 $ as $t \to 0$ and $l(t) \to k$ as $t \to 1$. Indeed, at the beginning, all the particles start at the origin and
stay next to $x=0$ at short times. At the end, the particle are split into $k$ groups, each composed of $n_i$ particles,
lying next to the $k$ and points $a_i(1)$. As time grows, the unique interval corresponding to $t=0$ splits into two intervals
for a critical value of time corresponding to a phase transition. Keeping time growing, these intervals split also for critical
values of time and this process keeps on iteratively until one gets $k$ intervals reaching the end points. In these notes, one
describes the distribution of Brownian particles in these intervals for any non critical value of time by studying the associated
matrix model for different regimes. In section \ref{seclarge}, we study the case of large time when $l(t)=k$ as a warm up
before studying the generic non-critical case $1\leq l(t)\leq k$ in section \ref{secinter}.

In both section, we follow the same procedure. We first study the Pastur equation \cite{Pastur}, or spectral curve,
\beq\label{eqpastur1}
\prod_{i=1}^k (y-a_i(t)) \left(x-y-\sum_{j=1}^k {n_j \over N(y-a_j(t))} \right) =0,
\eeq
of the matrix model obtained for time $t$, as well as the associated Riemann surface. The structure of the Riemann surface
changes with time: especially, it has as many cuts as the number $l$ of intervals $\left[z_{2i-1},z_{2i}\right]_{i=1}^l$ filled by the eigenvalues.
We then study an associated Riemann-Hilbert problem of size $(k+1) \times (k+1)$ whose solution $Y(x)$ allows to define a fundamental
kernel $K_N(x,y)$ giving access to any correlation function of the eigenvalues through:
\beq
R_m(x_1,x_2, \dots,x_m) = \det\left(K_N(x_j,x_k)\right)_{1\leq j,k \leq m}.
\eeq
By successive transformations of this RH problem, we turn it to a simple one exponentially close when $N \to \infty$ to a
model RH problem whose solution can be written explicitly.
This asymptotic solution of the RH problem leads to the main theorems of this paper. The first one describe the limiting density
of Brownian motion at a given time when we have a large number of them.
\bt\label{thmeandensity1}
The limiting density of eigenvalues
\beq
\rho(x):= \lim_{N\to \infty} {K_N(x,x)\over N}
\eeq
exists, is supported by the real cuts $\left[z_{2i-1},z_{2i}\right]_{i=1}^l$ and is expressed in terms of the unique solution
$y_0(x)$ of the Pastur's equation
\beq
\prod_{i=1}^k (y-a_i(t)) \left(x-y-\sum_{j=1}^k {n_j \over N(y-a_j(t))} \right) =0
\eeq
such that
\beq
y_0(x) \sim {1 \over x} + O({1 \over x^2}) \qquad \hbox{as} \qquad x \to \infty
\eeq
by
\beq
\rho(x) = {1 \over \pi} \left|\Im y_0(x) \right|.
\eeq
It is analytic on the cuts and vanishes as a square root at the edges $z_i$.
\et
The second one describes the behavior of the kernel inside the support of the eigenvalues (or Brownian motions).
\bt\label{thkernelbulk}
For any $x \in {\displaystyle \bigcup_{i=1}^{l}} (z_{2i-1},z_{2i})$ and any pair $(u,v) \in \mathbb{R}^2$, the kernel
converges to the universal sine kernel:
\beq\encadremath{
\lim_{N \to \infty} {1 \over N \rho(x)} \widehat{K}_N\left(x+{u \over N \rho(x)},x+{v \over N \rho(x)}\right) = {\sin \pi (u-v) \over \pi (u-v)}
}\eeq
where we use the rescaled kernel:
\beq
\widehat{K}_N(x,y) = e^{N(h(x)-h(y))} K_N(x,y)
\eeq
with the function
\beq
h(x) := -{x^2 \over 4} + \Re \left[\int^x y_0(x)\right]
\eeq
where the primitive will be made precise later.
\et
The last one describes the asymptotic behavior of the kernel at the edge of the support of eigenvalues $z_i$.
\bt\label{thkerneledge}
For any $i = 1, \dots, 2k$ and every $(u,v) \in \mathbb{R}^2$, the rescaled kernel is given by the universal Airy
kernel on the edges of the supports of eigenvalues:
\beq
\lim_{N \to \infty} {1 \over \left(\rho_i N\right)^{2 \over 3}} \widehat{K}_N\left( z_i + (-1)^i {u \over \left(\rho_i N\right)^{2 \over 3}},
z_i + (-1)^i {v \over \left(\rho_i N\right)^{2 \over 3}}\right) = {Ai(u) Ai'(v) - Ai'(u) Ai(v) \over u-v},
\eeq
where $\rho_i$ is a constant characterizing the behavior of $\rho$ near the branch point $z_i$:
\beq
\rho(x) = {\rho_i \over \pi} |x-z_i|^{1 \over 2} \left( 1 + o(1)\right)
\eeq
when $x \to z_i$.
\et


{\bf Organization of the paper}:
\begin{itemize}
\item section \ref{secgaussian} is a reminder on the Gaussian matrix model with an external field and its link to
non-intersecting Brownian motions.

\item section \ref{seclarge} is a warm up where we study the asymptotics of the kernel and mean density of eigenvalues in a
particular non degenerate case: there are as many supports of eigenvalues as the number of different eigenvalues of the
external matrix $A$: $l=k$.

\item in section \ref{secinter}, we study the general case $1\leq l \leq k$ and prove the theorems \ref{thmeandensity1},
\ref{thkernelbulk} and \ref{thkerneledge} for any non-critical spectral curve \eq{eqpastur1}.

\item section \ref{secconclu} is a conclusion and present some possible generalizations of the present work.

\end{itemize}

\section{Gaussian matrix integral in an external field, reminder}
\label{secgaussian}

\subsection{Definitions}

In this section we consider the hermitian matrix model defined through the partition function
\beq
Z:= \int_{H_N} dM e^{-N \Tr \left({M^2 \over 2} - A M \right)}
\eeq
where one integrates over the hermitian matrices $M$ of size $N \times N$, $A$ is a deterministic diagonal matrix of the form
\beq
A := \hbox{diag}(\overbrace{a_1, \dots,a_1}^{n_1},\overbrace{a_2,\dots,a_2}^{n_2}, \dots ,\overbrace{a_k,\dots,a_k}^{n_k})
\eeq
and the measure $dM$ is the product of the Lebesgue measure of the real entries of the matrix M:
\beq
dM := {\displaystyle \prod_{i<j}} d\Re\left(M_{ij}\right) d\Im\left(M_{ij}\right) {\displaystyle \prod_i} d M_{ii}.
\eeq

By diagonalizing the matrix $M$, we can reduce this problem to the study of its eigenvalues $x_1, \dots, x_N$, which are constrained by the
probability measure:
\beq
d\mu(x_1,\dots, x_N) = \prod_i dx_i {\Delta(x)\over \Delta(a_i)} e^{-N \sum_i {x_i^2 \over 2}} \det\left(e^{-x_i a_j}\right)
\eeq
where $\Delta(x)$ denotes the Vandermonde determinant and one considers the eigenvalues $a_i$ with their multiplicities.

They can be seen as a Coulomb gas composed $k$ groups of $n_k$ particles living on the real axis and submitted to a gaussian potential $V_i(x) = {x^2 \over 2} - a_i x$ and
a logarithmic repulsive interaction $\ln|x_i-x_j|$. The position of the eigenvalues is then an equilibrium configuration
for such particles. When the size of the matrix tends to infinity, the number of eigenvalues grows in the same way
as the latter and they fill a finite set of intervals $\left[z_{2i-1}, z_{2i}\right]_{i=1}^l$ with $1\leq l\leq k$. The number $l$ of such
intervals as well as their positions depend on the $a_i$'s and the ratios $\epsilon_i := {n_i \over N}$. They can
be encoded in an algebraic curve as we remind now.

Let us define the density of states as the one point correlation function:
\beq
\rho(x) := {1 \over N} \int_{\mathbb{R}^N} d\mu(u_1,u_2, \dots ,u_N) \sum_{i=1}^N \delta(x-u_i),
\eeq
as well as its Stieltjes transform, called the resolvent
\beq
W(x) := \int_{\mathbb{R}} {\rho(u) \over x-u} du.
\eeq

Remark that the density of states $\rho$ is expected, by the analogy with a Coulomb gas, to have a compact support composed of the
$l$ intervals corresponding to the $l$ saddles of the effective potential for the Coulomb particles. Thus, the
resolvent can be written:
\beq
W(x) = \sum_{i=1}^l {1 \over N} \int_{[z_{2i-1},z_{2i}]} {\rho(u) \over x-u} du
\eeq
and is continuous except cuts on the support $\cup_i [z_{2i-1},z_{2i}]$ of the density $\rho$.

One can show \cite{Pastur,Kuijlaars1,EOinvariants} that this resolvent satisfies an algebraic equation:
\beq
{\cal{E}}(x,W(x))=0
\eeq
where ${\cal{E}}(x,y)$ is  a polynomial in its variables. Once the so-called spectral curve ${\cal{E}}(x,y)=0$ is known, one thus obtain
$W(x)$ by solving a polynomial equation. Nevertheless, generically there exists as many solutions to this equation as the degree of this
polynomial in $y$. But, for some particular values of $x$, this equation has double roots in $y$. These particular points
$z_i$ are usually called branch points since they correspond to the points where two different solution of the equation meet.

It is easily seen that these branch points correspond to the $z_i$ fixing the endpoints of the different component of the
support of the density $\rho$. Finding the position of these branch points allows thus to know the support of the eigenvalues.

When the size of the matrices tends to infinity, the eigenvalues fill this support continuously and one is interested in
the behavior of the level density inside these support and on their boundaries. This is why we now study this spectral curve
in more details.

\subsection{Algebraic curve}

Following \cite{EOinvariants}, one can compute the spectral curve associated to the considered model.
For this model and arbitrary values of the $a_i$'s: the algebraic equation takes the form:
\beq\label{spectralinconnu}
{\cal{E}}(x,y) := (x-y) \prod_{i=1}^k (y-a_i) + P(y) = 0
\eeq
where $P(y)$ is a polynomial of degree $k-1$ whose coefficients must be fixed by the behavior of the function $y$ at infinity.

Indeed, this algebraic equation can be seen as the realization of a compact Riemann surface $\Sigma$ embedded in $\mathbb{CP}^1 \times \mathbb{CP}^1$
with two meromorphic functions $x$ and $y$ such that
\beq
\forall p \in \Sigma \,\, , \; \; {\cal{E}}(x(p),y(p))=0.
\eeq

One can easily see that this is a genus 0 curve composed of $k+1$-sheets in $y$. It means that for a given value of
$x(p) \in \mathbb{CP}^1$, there exist $k+1$ points on $\Sigma$ which are pre-images of $x$. We note them $p^0, p^1, \dots, p^k$:
\beq
\forall i=0,\dots , k \, , \; x(p^i) = x(p).
\eeq

To agree with the notations of \cite{Kuijlaars1,Kuijlaars2,Kuijlaars3}, one can define the value of the function $y$ on the different
sheets as $k+1$ different functions on the complex plane of the image $x(p)$:

\bd
Let the functions $\xi_i: \, \mathbb{C} \rightarrow \mathbb{C}$ be defined by
\beq
\xi_i(x(p)):= y(p^i).
\eeq
\ed

The different sheets can be discriminated by the behavior of $y(p)$ when $x(p) \to \infty$. From the definition, one can
label the different sheets in such a way that:
\beq
\begin{array}{l}
y(p^0) \sim_{x(p) \to \infty} x(p) + {1 \over x(p)} + O\left({1 \over x(p)^2}\right) \cr
\forall i \neq 0 \, , \; y(p^i) \sim_{x(p) \to \infty} a_i + {n_i \over N} {1 \over x(p)} + O\left({1 \over x(p)^2}\right),
\end{array}
\eeq
i.e. the $\xi$-functions have the following asymptotics:
\beq\label{asymptxi}
\begin{array}{l}
\xi_0(x) {\displaystyle \sim_{x \to \infty}} x + {1 \over x} + O\left({1 \over x^2}\right) \cr
\forall i \neq 0 \, , \; \xi_i(x) \sim_{x \to \infty} a_i + {n_i \over N} {1 \over x} + O\left({1 \over x^2}\right).
\end{array}
\eeq

Inserting these asymptotics inside the algebraic equation \eq{spectralinconnu}, one gets the coefficients of the polynomial
$P(y)$ and the algebraic curve reads:
\beq\label{spectralequation}\encadremath{
{\cal{E}}(x,y) = \prod_{i=1}^k (y-a_i) \left(x-y-\sum_{j=1}^k {n_j \over N(y-a_j)} \right).}
\eeq

\subsection{Kernel and correlation functions}

In \cite{KD1,KD2}, Daems and Kuijlaars extended classical results of random matrix theory to the case of 1 matrix model in an external
field: they explain how the correlation functions can be computed thanks to a unique kernel which can be linked to the solution
of an associate Riemann-Hilbert problem.

They showed that, using some multiple orthogonal polynomials, one can express the correlation functions of the model (see \cite{KD1,KD2,PZ1,PZ2}).
In particular, the joint probability density on the eigenvalues can be written
\beq
{1 \over N!} \det ( K_N(x_j,x_k))_{1\leq j,k\leq N }
\eeq
and the $m$-point correlation functions
\beq
R_m(x_1, \dots, x_m) = \det ( K_N(x_j,x_k))_{1\leq j,k\leq m }
\eeq
where the kernel $K_N(x,y)$ is expressed through some multi-orthogonal polynomials $\psi$ and $\phi$ by
\beq
K_N(x,y) = e^{-{1 \over 2} (x^2+y^2)} \sum_{k=0}^{N-1} \psi_k(x) \phi_k(y).
\eeq

Moreover, in \cite{KD1,KD2}, it was shown that these kernels associated to multiple orthogonal polynomials are solution of a general Riemann
Hilbert problem. Let us specify here the problem whose solution is the kernel giving correlation functions of the hermitian matrix
model introduced above with an external matrix $A$ whose eigenvalues are $a_i$, $i=1, \dots,k$.

\bt (Daems-Kuijlaars)

The correlation functions are given by
\beq
R_m(x_1, \dots, x_m) = \det ( K_N(x_j,x_k))_{1\leq j,k\leq m }
\eeq
where the kernel $K_N(x,y)$ reads:
\beq
K_N(x,y) = {1 \over 2 i \pi (x-y)} \Omega_1(y) Y_+^{-1}(y) Y_+(x) \Omega_2^t(x)
\eeq
where we define the vectors of size $k+1$:
\beq\label{defOmega1}
\Omega_1(x) = [0  \; e^{N{x^2 \over 4}} \omega_1(x) \; \dots \; e^{N{x^2 \over 4}} \omega_k(x)]
\eeq
with $\omega_i := e^{-N({x^2 \over 2} - a_i x)}$ and
\beq\label{defOmega2}
\Omega_2(x) = [1 \; 0 \; \dots \; 0],
\eeq
and $Y(x)$ is the unique solution of the following Riemann Hilbert problem:

\begin{itemize}

\item $Y: \mathbb{C} \; \backslash \; \mathbb{R} \to \mathbb{C}^{(k+1) \times (k+1)}$ is analytic;

\item for $x \in \mathbb{R}$, one has the jumps
\beq
Y_+(x) = Y_-(x) \left( \begin{array}{ccccc}
1 & \omega_1(x) & \omega_2(x) & \dots & \omega_k(x) \cr
0 & 1 & 0 & \dots & 0 \cr
0 & 0 & 1 & \dots & 0 \cr
\dots & \dots & \dots & \dots & \dots \cr
0 & 0 & 0 & \dots & 1 \cr
\end{array} \right)
\eeq
where  $Y_+(x)$ and $Y_-(x)$ denote respectively the limit of $Y(z)$ when
$z \to  x$ from the upper and lower half planes;

\item when $x \to \infty$, one has the asymptotic behavior:
\beq
Y(x) = \left({\bf I} + O\left( {1 \over x}\right)\right) \left(
\begin{array}{ccccc}
z^N & 0 & 0 & \dots & 0 \cr
0 & z^{-n_1} & 0 & \dots & 0 \cr
0 & 0 & z^{-n_2} & \dots & 0 \cr
\dots & \dots & \dots & \dots & \dots \cr
0 & 0 & 0 & \dots & z^{-n_k} \cr
\end{array} \right)
\eeq
where ${\bf I}$ denotes the identity matrix of size $(k+1) \times (k+1)$.

\end{itemize}

\et

\subsection{Non-intersecting Brownian motions}

One of the main motivations for this work is the study of a physical system composed of many $N$ Brownian motions as $N$
gets large. Let us consider $N$ Brownian motions $x_i(t)$, $i=1, \dots ,N$,  on the real axis starting from the origin
at time $t=0$:
\beq
\forall i = 1, \dots, N \, , \qquad x_i(0)=0.
\eeq
Let us impose that they cannot intersect and that their final positions at time $t$ is given by a set of $k$ ordered distinct real points
$a_i$:
\beq
\left(x_1(1),x_2(1),x_3(1), \dots , x_N(1) \right)=
\left(\overbrace{a_1, \dots,a_1}^{n_1},\overbrace{a_2,\dots,a_2}^{n_2}, \dots ,\overbrace{a_k,\dots,a_k}^{n_k}\right)
\eeq
for any integers $\left\{n_i\right\}_{i=1}^k$ such that
\beq
\sum_{i=1}^k n_i = N.
\eeq
The physical quantities describing this statistical system at a given time $0<t<1$ are the correlation functions:
\beq
R_m(X_1,X_2, \dots, X_M;t) := {N! \over (N-m)!} \int_{\mathbb{R}^{N-m}} dP(x_1(t),x_2(t),\dots,x_N(t))
\prod_{i=1}^m \delta(x_i-X_i)
\eeq
where $P(x_1(t),x_2(t), \dots, x_N(t))$ is the probability measure on the Brownian motions at time $t$.

This problem does not look easy to solve at first sight. Fortunately, there exists an underlying integrable structure
and it was proved that these correlation functions coincide with the correlation functions of a hermitian gaussian matrix model
submitted to an external matrix
\beq
A:= A(t) :=
\hbox{diag}(\overbrace{a_1(t), \dots,a_1(t)}^{n_1},\overbrace{a_2(t),\dots,a_2(t)}^{n_2}, \dots ,\overbrace{a_k(t),\dots,a_k(t)}^{n_k})
\eeq
where the eigenvalues $a_i$ are just rescaled versions of the ending points:
\beq
a_i(t):= a_i \sqrt{t\over 1-t}.
\eeq
In order to compute these correlation functions at time $t$, one is then left with the study of the corresponding random
matrix $M(t)$ whose statistic depends on $t$ only through the external matrix eigenvalues $a_i(t)$. Remark also that the time
evolution of these external eigenvalues is very simple.

Let us now interpret the ingredients of the matrix model's study in this statistical physics setup to get some intuition.
First of all, one sees that the eigenvalues of the random matrix $x_i(t)$ are identified with the position of the
Brownian motions at time $t$. The cuts of the spectral curve $\left[z_{2i-1}(t),z_{2i}(t)\right]_{i=1}^l$ should then coincide
with the support of the Brownian motions at time $t$. One can then study the evolution of these support by studying the
time evolution of their edges, i.e. the time evolution of the branch points.

Consider the equation of the algebraic curve \eq{spectralequation}. The inversion formula giving $x$ in terms of $y$ reads\footnote{Remark that
this gives a natural rational parametrization of this curve: \beq \left\{\begin{array}{l} y=z \cr x= z + \sum_{i=1}^k {n_i \over N (z-a_i(t))}\cr
\end{array}\right.\eeq}
\beq\encadremath{
x= y + \sum_{i=1}^k {n_i \over N (y-a_i(t))}}
\eeq
and has, generically,  $2k$ branch points given by the roots of the equation:
\beq\label{brancheq}
1 = \sum_{i=1}^k {n_j \over N (y-a_j(t))^2} .
\eeq
Let us call $\{z_i(t)\}_{i=1}^{2k}$ these branch points ordered following their real part.
Since the $a_j$'s are real, the roots of \eq{brancheq} are either real or come in pairs of complex conjugated points.

Let us now have a look at the two extremal cases corresponding to the boundaries. For $t\to 1$, the eigenvalues diverge
$a_i(t)\to \infty$. Thus, the equation for the branch points \eq{brancheq} as solutions:
\beq
y \sim a_i(1) \sim \infty,
\eeq
i.e. one has $2k$ real branch points of the type $z_{2i-1}<a_i<z_{2i}$ with $z_{2i-1}$ and $z_{2i}$  very close to $a_i$. This
confirms the physical intuition of the behavior of the Brownian motions: they merge in $k$ groups close to the end points
$a_i$ when $t \to 1$.
On the contrary, as $t \to 0$, all the $a_i(t)$ merge to $0$ and all the branch points merge also to 0: only two
of them are real, the other $2k-2$ come in pairs of complex conjugated points.

Let us now come to the generic configuration. Starting from $t=1$, as the time decreases the real branch points $z_{2i-1}$
and $z_{2i}$ move away from $a_i$. When $z_{2i}$ and $z_{2i+1}$ meet, one reaches a critical point and the corresponding configuration is not
studied in this paper but will be addressed elsewhere. After crossing this phase transition, these two branch points move
away from the real axis becoming complex conjugated. Then, at any intermediate time, a generic configuration consists in
$2l$ real branch points $(z_{2i-2},z_{2i})$, $i=1,\dots,l$, where the number of real cuts $l$ decreases with time from
$k$ to $1$\footnote{The number $l$ corresponds to the number of disjoint segments filled by the Brownian motions
at a given time $t$. The proof of this property is not technically involved but rather long, that is why it was
decided not to include it in the present paper since the interested reader can easily derive it.}. The other branch points are complex conjugated numbers $\left(w_i^{(h)},\overline{w}_i^{(h)}\right)$
whose real part belong to the segments $\left[z_{2i-2},z_{2i}\right]$.

In the following sections, we study the Riemann surface associated to these different configuration: in section \ref{seclarge},
we consider the case where we have only real branch points whereas in section \ref{secinter}, we study the generic case
with an arbitrary number of real eigenvalues.

\section{Large time case}
\label{seclarge}

In this section, we assume that all the branch points are real. This configuration is encountered for large times in the brownian
motion setup: it corresponds to having all the particles in the neighborhood of their end points.
Then one can simply mimic the computations of Bleher and Kuijlaars \cite{Kuijlaars1} and study the algebraic curve.

\subsection{Study of the spectral curve}\label{spectsimple}

Before solving the Riemann-Hilbert problem, one needs to study the associated spectral curve more carefully.
Remember that, in the large time case we consider in this paragraph, the spectral curve has $2k$ real branch points
$\{z_i\}_{i=1}^{2k}$.

We can show the following lemmas concerning the sheet structure of the spectral curve built out of the $y$ function.
First of all, we can build analytical continuations of the $\xi_i$ as follows:
\bl
The functions $\xi_i$ for $i \neq 0$ (resp. $\xi_0$) are analytic in
$\mathbb{C}\backslash [z_{2i-1},z_{2i}]$ (resp. $\mathbb{C}\backslash {\displaystyle \bigcup_{j=1}^k} [z_{2j-1},z_{2j}]$).
\el

\proof{
It is easily proved by starting from the asymptotics \eq{asymptxi} and analytically continuing these functions using the rationnal
parametrization of this genus 0 curve.}

These functions satisfy the following jump conditions:
\bl
For any $i = 1, \dots, k$ and $x \in [z_{2i-1},z_{2i}]$
\beq
\xi_{0+}(x) = \overline{\xi_{0-}(x)} = \xi_{i-}(x) = \overline{\xi_{i+}(x)}
\eeq
and for $x\in \bigcup_{i=1}^k ]z_{2i-1},z_{2i}[$
\beq
\Im \xi_{0+}(x) >0
\eeq
\el
\proof{It also follows directly from the definitions.
}

\begin{figure}
  \hspace{4cm} \includegraphics[width=8cm]{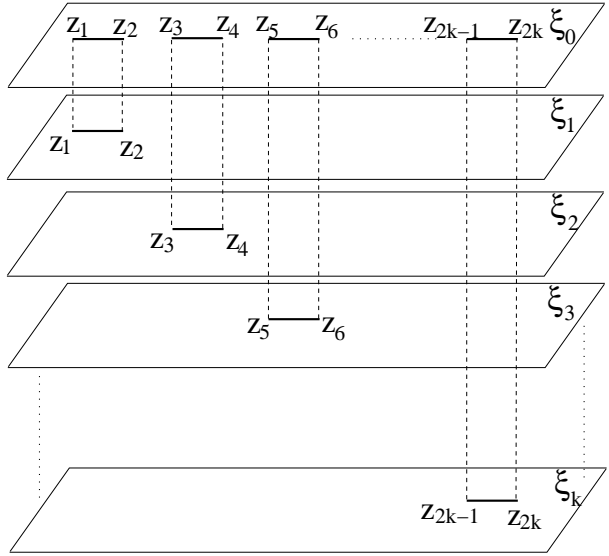}\\
  \caption{Sheeted structure of the spectral curve in the large time case. The sheet 0 is linked to the other $k$ sheets
by non intersecting segments.}\label{figlarge}
\end{figure}

It means that the spectral curve is composed of $k+1$ sheets linked by $k$ real cuts as follows: the sheet $i$ is linked to
the sheet 0 by the cut $[z_{2i-1},z_{2i}]$ (see fig.\ref{figlarge}).

We can then define the "eigenvalue density" corresponding to the density of brownian particles at a given point $x$ of the
real axis:
\bd
Let us define, for any $x \in \bigcup_{j=1}^k [z_{2j-1},z_{2j}]$,
\beq
\rho(x):= {1 \over \pi} \Im \xi_{0+}(x).
\eeq
\ed

One can now study its properties giving the physically expected result:
\bl\label{lemedge}
For $x\in {\displaystyle \bigcup_{i=1}^k} ]z_{2i-1},z_{2i}[$, $\rho(x)>0$ and
\beq
\int_{z_{2i-1}}^{z_{2i}} \rho(x) dx = {n_i \over N}.
\eeq

On the edges of the spectrum, there exists $\rho_i>0$ such that
\beq
\rho(x) = {\rho_i \over \pi} |x-z_i|(1 + O(z-z_i)) \; \; \hbox{as} \;\; x \to z_i .
\eeq
\el

\proof{From the preceding lemma, one directly get that
\beq
\forall x\in \bigcup_{i=1}^k ]z_{2i-1},z_{2i}[ \, ,\;  \rho(x)>0 .
\eeq

Consider now any integer $i \in [1,k]$ and $x \in ]z_{2i-1},z_{2i}[$, one has
\beq
\rho(x) = {1 \over \pi} \Im \xi_{0+}(x) = {1 \over 2 i \pi} (\xi_{0+}(x) - \overline{\xi_{0+}(x)}) = {1 \over 2 i \pi } (\xi_{i-}(x) - \xi_{i+}(x))
\eeq
and thus, one can compute the integral
\beq
\int_{z_{2i-1}}^{z_{2i}} \rho(x) = {1 \over 2 i \pi} \int_{z_{2i-1}}^{z_{2i}} (\xi_{i-}(x) - \xi_{i+}(x))
= {1 \over 2 i \pi} \oint_{\Gamma_i} \xi_{i}(x) dx
\eeq
where the integration contour $\Gamma_i$ encircles the cut $[z_{2i-1},z_{2i}]$. Since $\xi_{i}$ has no other poles except
infinity, one can move it and use the asymptotics:
\beq
\int_{z_{2i-1}}^{z_{2i}} \rho(x) dx = - \Res_{x \to \infty} \xi_{i}(x) dx = {n_i \over N}.
\eeq

The last part of the lemma follows from the usual behavior on the edge of the spectrum (see for example \cite{Kuijlaars1}).
}

\br
All the results of this lemma simply follows on the behavior of he 1-form $ydx$ in the neighborhood of a cut linking two simple
branch points. Thus all these {\em local} result could be obtained by the former study of simpler models where one has only two sheets
for example.
\er

We now define the integrals of the $\xi_i$ functions as:
\bd
We define the integrals $\l_k$ by
\beq
\l_k(z):= \int^z \xi_k(x) dx
\eeq
where the integration constants are constrained by:
\beq\label{constr0large}
\l_0(z_{2k}) = \l_k(z_{2k}) = 0
\eeq
and
\beq\label{constrilarge}
\forall i = 1, \dots , k-1 \, , \;  \l_i(z_{2i}) = \l_{0+}(z_{2i}).
\eeq
\ed

\bl
With this definition, the jumps and asymptotics of these $\l$-functions are given by:
\begin{itemize}
\item When $z \to \infty$:
\beq
\l_0(z) = {z^2\over 2} - \ln z + l_0 + O\left({1 \over z^2}\right)
\eeq
and, for $i \neq 0$:
\beq
\l_i(z) = a_i z + {n_i \over N} \ln z + l_i + O\left({1 \over z^2}\right)
\eeq
where $l_i$'s are some constants chosen to satisfy the constraints \eq{constr0large} and \eq{constrilarge}.

\item They follow the jumps:
\beq
\begin{array}{ll}
\l_{0+}(x) - \l_{0-}(x) = - 2 i \pi \sum_{j=i+1}^k {n_i \over N}  & x \in [z_{2i},z_{2i+1}] \cr
\l_{0+}(x) - \l_{0-}(x) = - 2 i \pi & x \in (-\infty,z_1] \cr
\forall i \in [1,k] \, , \; \l_{i+}(x) - \l_{i-}(x) = 2 i \pi {n_i \over N} & x \in (-\infty,z_{2i-1}] \cr
\l_{0+}(x) - \l_{k-}(x) = 0 \, , \; \l_{0-}(x) - \l_{k+}(x) = 0 & x \in [z_{2k-1},z_{2k}] \cr
\forall i \in [1,k-1] \, , \; \l_{0+}(x) - \l_{i-}(x) = 0 \, , \; \l_{0-}(x) - \l_{i+}(x) = 2 i \pi \sum_{j=i+1}^k {n_i \over N} & x \in [z_{2i-1},z_{2i}]
\end{array}
\eeq
\end{itemize}
\el

\proof{
The first part of this lemma is just the rewriting of the integration of the asymptotics of $ydx$ on the different sheets.

The second part is obtained by moving the integration contours so that one is left by the computation of a residue at infinity
on the different sheets.
}

Let us now state a fundamental theorem for the following
\bt\label{thneigh}
For any $i= 1, \dots, k$, on $\mathbb{R}\backslash[z_{2i-1},z_{2i}]$, one has $\Re \l_{i+} < \Re \l_{0-}$.

For any $i= 1, \dots, k$, there exists a neighborhood $U_i$ of $]z_{2i-1},z_{2i}[$ such that
\beq
\forall j \neq i \, , \; \Re \l_j(z) < \Re \l_0(z) < \Re \l_i(z)
\eeq
for any $z \in U_i-]z_{2i-1},z_{2i}[$.
\et
\proof{
Once again, one can follow \cite{Kuijlaars1} for the first part of the theorem:
for $x>z_{2i}$, $\xi_0(x) > \xi_i(x)$, and $\l_0(x)>\l_i(x)$ because $\l_i(z_{2i}) = \l_0(z_{2i})$ and $\l_i'= \xi_i$;
for $x<z_{2i-1}$, one easily sees that $\Re \xi_o(x) < \xi_i(x)$ which gives the result.

For the second part, one has to study the properties of the $\l$-functions more carefully.
Consider the function $F_i := \l_{i+}- \l_{0+}$. It is purely imaginary on $]z_{2i-1},z_{2i}[$ and its derivative
$F'(x) = -2i \pi \rho(x)$ has negative imaginary part. Thus $\Re \l_0(z) < \Re \l_i(z)$ in a neighborhood of $]z_{2i-1},z_{2i}[$.
The other inequality follows from the definition and continuity.
}

\subsection{Riemann-Hilbert analysis}

In this section, we solve the Riemann-Hilbert problem leading to the computation of the kernel. We look for the unique function
$Y$ defined as follows:
\bd
Let $Y$ be the solution of the following constraints:
\begin{itemize}

\item $Y \, : \, \mathbb{C} \backslash \mathbb{R} \rightarrow \mathbb{C}^{(k+1) \times (k+1)}$ is analytic;

\item for $x \in \mathbb{R}$, $Y(x)$ has the jumps
\beq
Y_+(x) = Y_-(x) \left( \begin{array}{ccccc}
1 & \omega_1(x) & \omega_2(x) & \dots & \omega_k(x) \cr
0 & 1 & 0 & \dots & 0 \cr
0 & 0 & 1 & \dots & 0 \cr
\dots & \dots & \dots & \dots & \dots \cr
0 & 0 & 0 & \dots & 1 \cr
\end{array} \right)
\eeq
where $\omega_i(x)  := e^{-N({x^2 \over 2} - a_i x)}$ and $Y_+(x)$ and $Y_-(x)$ denote respectively the limit of $Y(z)$ when
$z \to  x$ from the upper and lower half planes;

\item when $x \to \infty$, one has the asymptotic behavior:
\beq
Y(x) = \left({\bf I} + O\left( {1 \over x}\right)\right) \left(
\begin{array}{ccccc}
z^N & 0 & 0 & \dots & 0 \cr
0 & z^{-n_1} & 0 & \dots & 0 \cr
0 & 0 & z^{-n_2} & \dots & 0 \cr
\dots & \dots & \dots & \dots & \dots \cr
0 & 0 & 0 & \dots & z^{-n_k} \cr
\end{array} \right)
\eeq

\end{itemize}

\ed

The kernel can then be written:
\beq
K_N(x,y) = {1 \over 2 i \pi (x-y)} \Omega_1(y) Y_+^{-1}(y) Y_+(x) \Omega_2^t(x)
\eeq
with the notations of \eq{defOmega1} and \eq{defOmega2}.
To do so, we transform the problem step by step to turn it to a simple one (see \cite{Bleher} for a review on the subject).

It is important to keep in mind the constraints we impose on the $n_i'$'s and N:
\begin{itemize}
\item We fix a set of $k$ positive rational number called filling fractions:
\beq
\epsilon_i = {p_i\over q_i}
\eeq
where $p_i$ and $q_i$ are coprime and
\beq
\sum_{i=1}^k \epsilon_i = 1.
\eeq

\item $N$ is a common multiple of the $q_i$'s.

\item The integers $n_i$ are fixed by the condition:
\beq
n_i = \epsilon_i N.
\eeq

\end{itemize}
With these constraints, the spectral curve associated to this RH problem does not change as $N$ grows.

\subsubsection{First transformation}
Using the notations of section \ref{spectsimple}, we can define
\beq
T(x):= \hbox{diag}\left(e^{-N l_0},e^{-N l_1}, \dots, e^{-N l_k}\right) Y(x)
\hbox{diag}\left(e^{N \left(\l_0 - {x^2 \over 2}\right)},e^{N \left(\l_1 - {n_1 \over N} x\right)},e^{N \left(\l_2 - {n_2 \over N} x\right)}, \dots, e^{N \left(\l_k - {n_k \over N} x\right)}\right)
\eeq
for $x \in \mathbb{C} \backslash \mathbb{R}$.

Then one can check that it is solution of the Riemann-Hilbert problem:
\begin{itemize}
\item $T$ is analytic on $\mathbb{C} \backslash \mathbb{R}$;

\item for $x \in \mathbb{R}$, one has the jumps
\beq
T_+(x) = T_-(x) j_T(x)
\eeq
where
\beq
j_T(x) = \left( \begin{array}{ccccc}
e^{N(\l_{0+}(x)-\l_{0-}(x))} & e^{N(\l_{1+}(x)-\l_{0-}(x))} & e^{N(\l_{2+}(x)-\l_{0-}(x))} & \dots & e^{N(\l_{k+}(x)-\l_{0-}(x))} \cr
0 & e^{N(\l_{1+}(x)-\l_{1-}(x))} & 0 & \dots & 0 \cr
0 & 0 & e^{N(\l_{2+}(x)-\l_{2-}(x))} & \dots & 0 \cr
\dots & \dots & \dots & \dots & \dots \cr
0 & 0 & 0 & \dots & e^{N(\l_{k+}(x)-\l_{k-}(x))} \cr
\end{array} \right).
\eeq

\item when $x \to \infty$, one has the asymptotic behavior:
\beq
T(x) = \left({\bf I} + O\left( {1 \over x}\right)\right).
\eeq

\end{itemize}

Note that one can write the jump matrix $j_T(x)$ depending on the cut on which $x$ lies:
\vs

For $x \in [z_{2i-1},z_{2i}]$, $j_T$ takes the form
{\tiny \beq
\left( \begin{array}{cccccccc}
e^{N(\l_{0}(x)-\l_{i}(x))_+} & e^{N(\l_{1+}(x)-\l_{0-}(x))} & \dots & e^{N(\l_{i-1+}(x)-\l_{0-}(x))}& 1 & e^{N(\l_{i-1+}(x)-\l_{0-}(x))}& \dots & e^{N(\l_{k+}(x)-\l_{0-}(x))} \cr
0 & 1 & \dots & 0 & 0 & 0 & \dots & 0 \cr
\dots &  \dots  & \dots &  \dots  &  \dots  &  \dots  & \dots &  \dots  \cr
0 &  0 & \dots & 1 & 0 & 0 & \dots & 0 \cr
0 &  0 & \dots & 0 & e^{N(\l_{0}(x)-\l_{i}(x))_-} & 0 & \dots & 0 \cr
0 &  0 & \dots & 0 & 0 & 1 & \dots & 0 \cr
\dots &  \dots  & \dots &  \dots  &  \dots  &  \dots  & \dots &  \dots  \cr
0 &  0 & \dots & 0 & 0 & 0 & \dots & 1 \cr
\end{array} \right).
\eeq}

For $x \in \mathbb{R} \backslash {\displaystyle \bigcup_{i}} [z_{2i-1},z_{2i}]$, $j_T$ reduces to
\beq
j_T = \left( \begin{array}{ccccc}
1 & e^{N(\l_{1+}(x)-\l_{0-}(x))} & e^{N(\l_{2+}(x)-\l_{0-}(x))} & \dots & e^{N(\l_{k+}(x)-\l_{0-}(x))} \cr
0 & 1 & 0 & \dots & 0 \cr
0 & 0 & 1 & \dots & 0 \cr
\dots & \dots & \dots & \dots & \dots \cr
0 & 0 & 0 & \dots & 1 \cr
\end{array} \right).
\eeq

The kernel can then be written:
\beq
K_N(x,y) = { e^{N\left({x^2 \over 4}-{y^2 \over 4}\right)}\over 2 i \pi (x-y)} \Omega_{1,T}(y) T_+^{-1}(y) T_+(x) \Omega_{2,T}^t(x)
\eeq
where
\beq
\Omega_{1,T}(x) = [0 \; e^{N \l_{1+}(x)} \; \dots \; e^{N \l_{k+}(x)}]
\eeq
and
\beq
\Omega_{2,T}(x) = [e^{-N \l_{0+}(x)} \; 0 \; \dots \; 0].
\eeq

\subsubsection{Second transformation and opening of lenses}

Remark hat one can further factorize the matrix $j_T(x)$ for $x \in [z_{2i-1},z_{2i}]$:
\beq
j_T(x) = \widetilde{j}_{i,T}(x)
j_S(x)
\widehat{j}_{i,T}(x)
\eeq
where, for $x \in [z_{2i-1},z_{2i}]$, one has defined the $k+1\times k+1$ matrices:
\beq
j_S(x) = \left[(j_S)_{\alpha \beta}\right]_{\alpha,\beta = 0}^k  := \left[(1-\delta_{\alpha i}-\delta_{\alpha 0})\delta_{\alpha \beta}
+ \delta_{\alpha 0} \delta_{\beta i} - \delta_{\alpha i} \delta_{\beta 0}\right]_{\alpha,\beta = 0}^k ,
\eeq
\beq
\widetilde{j}_{i,T}(x) := \left[ \delta_{\alpha \beta} +\delta_{\alpha i} (\delta_{\beta 0} e^{N(\l_0-\l_i)_-} -\sum_{j\neq 0,i} \delta_{\beta j} e^{N(\l_j-\l_i)_-})  \right]_{\alpha,\beta = 0}^k
\eeq
and
\beq
\widehat{j}_{i,T}(x) := \left[ \delta_{\alpha \beta} +\delta_{\alpha i} (\sum_{j\neq i} \delta_{\beta j} e^{N(\l_j-\l_i)_+})  \right]_{\alpha,\beta = 0}^k .
\eeq

\begin{figure}
\hspace{3cm}  \includegraphics[width=11cm]{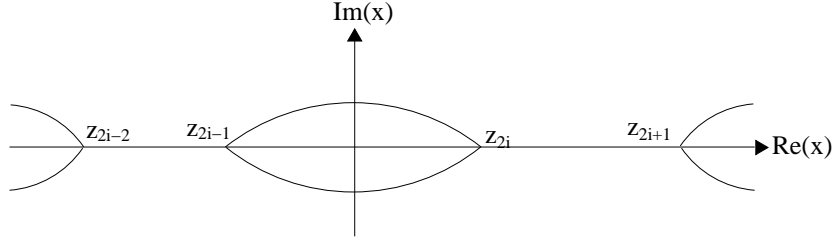}\\
  \caption{Lenses used in the first transformation: the lense i has vertices $z_{2i-1}$ and $z_{2i}$.}\label{lenselarge}
\end{figure}

This allows us to define the second transformation. Consider a set of $k$ lenses with vertices $z_{2i-1}$ and $z_{2i}$ (see fig.\ref{lenselarge}) contained in the
neighborhood $U_i$ defined in lemma \ref{thneigh} and define
\beq
S(z):= \left\{
\begin{array}{l}
T(z) \widehat{J}_i(x) \;\; \hbox{in the upper lens region} \cr
T(z) \widetilde{J}_i(x) \;\; \hbox{in the lower lens region} \cr
\end{array} \right.
\eeq
with
\beq
\widetilde{J}_{i}(x) := \left[ \delta_{\alpha \beta} +\delta_{\alpha i} (\delta_{\beta 0} e^{N(\l_0-\l_i)} -\sum_{j\neq 0,i} \delta_{\beta j} e^{N(\l_j-\l_i)})  \right]_{\alpha,\beta = 0}^k
\eeq
and
\beq
\widehat{J}_{i}(x) := \left[ \delta_{\alpha \beta} - \delta_{\alpha i} (\sum_{j\neq i} \delta_{\beta j} e^{N(\l_j-\l_i)})  \right]_{\alpha,\beta = 0}^k .
\eeq
for $z$ in the lens $i$ , and
\beq
S(z) := T(z)
\eeq
for $z$ outside the lenses.

We now define the jump matrix $j_S(x)$ on all the discontinuity contours of $S(x)$:
\begin{itemize}
\item On $[z_{2i-1},z_{2i}]$, for $i = 1, \dots, k$:
\beq
j_S(x) = \left[(1-\delta_{\alpha i}-\delta_{\alpha 0})\delta_{\alpha \beta}
+ \delta_{\alpha 0} \delta_{\beta i} - \delta_{\alpha i} \delta_{\beta 0}\right]_{\alpha,\beta = 0}^k ,
\eeq

\item On the upper boundary of the lens $i$:
\beq
j_S(x) = \left[ \delta_{\alpha \beta} - \delta_{\alpha i} (\sum_{j\neq i} \delta_{\beta j} e^{N(\l_j-\l_i)})  \right]_{\alpha,\beta = 0}^k ,
\eeq

\item On the lower boundary of the lens $i$:
\beq
j_S(x) = \left[ \delta_{\alpha \beta} +\delta_{\alpha i} (\delta_{\beta 0} e^{N(\l_0-\l_i)} -\sum_{j\neq 0,i} \delta_{\beta j} e^{N(\l_j-\l_i)})  \right]_{\alpha,\beta = 0}^k,
\eeq

\item On $\mathbb{R} \backslash {\displaystyle \bigcup_i} [z_{2i-1},z_{2i}]$ :
\beq
j_S(x) = j_T(x).
\eeq

\end{itemize}

Using these definition, one sees that $S(x)$ is solution of the following RH problem:
\begin{itemize}
\item $S(x)$ is analytic on $\mathbb{C} \backslash (\mathbb{R}\bigcup \Gamma)$ where $\Gamma$ denotes the boundaries of the lenses;

\item $S(x)$ has jumps:
\beq
\forall x \in \mathbb{R}\bigcup \Gamma \, , \; S_+(x) = S_-(x) j_S(x);
\eeq

\item when $x \to \infty$:
\beq
S(x) = {\bf I} + O\left({1 \over x}\right)
\eeq

\end{itemize}

and the kernel can be expressed in terms of this matrix $S$ by the following expression:

for $(x,y) \in (z_{2i-1},z_{2i})$

\beq\label{kernelS}
K_N(x,y) = { e^{N\left[h(y)-h(x)\right]}\over 2 i \pi (x-y)} \Omega_{1,S,i}(y) S_+^{-1}(y) S_+(x) \Omega_{2,S,i}^t(x)
\eeq
where
\beq
h(x) = -{x^2 \over 4} + \Re\left[\l_{0+}(x)\right],
\eeq
\beq
\Omega_{1,S,i}(x) = [ - e^{N i \Im\left[\l_{0+}(x)\right]} \; \overbrace{0 \; \dots \; 0}^{i-1} \; e^{-N i \Im\left[\l_{0+}(x)\right]} \;\overbrace{0 \; \dots \; 0}^{k-i}]
\eeq
and
\beq
\Omega_{2,S,i}(x) = [ e^{-N i \Im\left[\l_{0+}(x)\right]} \; \overbrace{0 \; \dots \; 0}^{i-1} \; e^{N i \Im\left[\l_{0+}(x)\right]} \;\overbrace{0 \; \dots \; 0}^{k-i}].
\eeq

One can see that, as $N \to \infty$, this RH problem reduces to a model RH of the same kind of that studied in \cite{Kuijlaars1,Kuijlaars2,Kuijlaars3}
thanks to the properties of the $\l$-functions in section \ref{spectsimple}. Let us solve it in the same way.

\subsubsection{Model RH problem.}
This means that we want to solve the following RH problem:
\begin{itemize}
\item $M(x): \mathbb{C}\backslash  {\displaystyle \bigcup_{i=1}^k} [z_{2i-1},z_{2i}] \rightarrow \mathbb{C}^{k+1 \times k+1}$;

\item $M(x)$ is analytic in $\mathbb{C}\backslash  {\displaystyle \bigcup_{i=1}^k} [z_{2i-1},z_{2i}]$;

\item $M$ has jumps on the cuts $[z_{2i-1},z_{2i}]$ given by:
\beq
\forall x \in \bigcup_{i=1}^k [z_{2i-1},z_{2i}] \, , \; M_+(x) = M_-(x) j_S(x);
\eeq

\item as $x \to \infty$:
\beq
M(x) = {\bf I} + O\left({1 \over x}\right).
\eeq

\end{itemize}

Let us lift this problem to the associated spectral curve by cutting the complex plane along the different $y$-sheets:
one defines:
\beq
\Omega_i = \xi_i(\mathbb{C}\backslash [z_{2i-1},z_{2i}])
\eeq
for $i = 1, \dots, k$ and
\beq
\Omega_0 = \xi_0(\mathbb{C}\backslash \bigcup_{i=1}^k [z_{2i-1},z_{2i}]).
\eeq
These images of the different $y$-sheets give a partition of the complex plane by definition and we can define the boundaries
$\Gamma_i = \Gamma_i^+ \bigcup \Gamma_i^-$
separating $\Omega_0$ and $\Omega_i$ for any $i\neq 0$ as follows:
\beq
\Gamma_i^+ = \Gamma_i \bigcap \{\Im z \geq 0\} = \xi_{0+}([z_{2i-1},z_{2i}])  = \xi_{i-}([z_{2i-1},z_{2i}])
\eeq
and
\beq
\Gamma_i^- = \Gamma_i \bigcap \{\Im z \leq 0\} = \xi_{0-}([z_{2i-1},z_{2i}])  = \xi_{i+}([z_{2i-1},z_{2i}]).
\eeq

We are looking for a solution under the form of a Lax matrix
\beq
M(x) = \left(
\begin{array}{cccc}
\phi_0(\xi_0(x)) &\phi_0(\xi_1(x)) & \dots & \phi_0(\xi_k(x)) \cr
\phi_1(\xi_0(x)) &\phi_1(\xi_1(x)) & \dots & \phi_1(\xi_k(x)) \cr
\dots & \dots & \dots& \dots \cr
\phi_k(\xi_0(x)) &\phi_k(\xi_1(x)) & \dots & \phi_k(\xi_k(x)) \cr
\end{array} \right)
\eeq
where $\phi$ is a Baker-Akhiezer type function analytic on $\mathbb{C} \backslash \bigcup_i \Gamma_i$ which must satisfy the
jump conditions:
\beq
\forall x \in \bigcup_j \Gamma_j^+ \, , \; \phi_{i+}(\xi) = \phi_{i-}(\xi)
\eeq
and
\beq
\forall x \in \bigcup_j \Gamma_j^- \, , \; \phi_{i+}(\xi) = - \phi_{i-}(\xi).
\eeq
These functions are also constrained by their asymptotic behavior:
\beq
\forall i,j = 0 , \dots , k \, , \; \phi_i(a_j) = \phi_i(\xi_j(\infty)):=\delta_{ij}
\eeq
where one conventionally notes $a_0 := \infty$.

One also notes
\beq
\Gamma_i \bigcap \mathbb{R} = \{p_{2i-1},p_{2i}\} =  \{\xi_i(z_{2i-1}),\xi_i(z_{2i})\}.
\eeq

One has the solution:
\beq
\phi_0(\xi) = {{\displaystyle \prod_{i=1}^k} (\xi- a_i) \over \sqrt{{\displaystyle \prod_{i=1}^k} (\xi-p_{2i-1})(\xi-p_{2i})}}
\eeq
and
\beq
\phi_i(x) = c_i {{\displaystyle \prod_{j \neq i}} (\xi - a_j) \over \sqrt{{\displaystyle \prod_{i=1}^k} (\xi-p_{2i-1})(\xi-p_{2i})}}
\eeq
for $i = 1, \dots, k$, where the $c_i$'s are constants which can be fixed by the asymptotic behavior of $\phi_i$'s and the square roots
have branch cuts along $\bigcup \Gamma_i$.

Indeed, the algebraic equation defining the branch points reads:
\beq
\prod_{i=1}^k (\xi-p_{2i-1})(\xi-p_{2i}) = {\displaystyle \prod_{i=1}^k} (\xi-a_i)^2 - \sum_{i=1}^k {n_i \over N} \prod_{j \neq i} \left(\xi-a_j \right)^2
\eeq
which gives
\beq
\phi_i(a_i) = c_i \sqrt{-{N \over n_i}}
\eeq
i.e.
\beq
c_i = - i \sqrt{n_i \over N}.
\eeq

\subsubsection{Parametrix at the edge points}

Now the solution of the RH problem can be build away from the branch points thanks to the preceding section, let us have a closer
look at the branch points $z_j$. Since the behavior near the branch points is the typical square root
and one can use the usual method (see for example \cite{Kuijlaars1}).

The problem is to find a parametrix in the neighborhood of the branch points, i.e. a solution to the RH problem:
\begin{itemize}
\item $P(z)$ is analytic in ${\displaystyle \bigcup_{i=1}^{2k}} D(z_i,r)  \backslash (\mathbb{R} \cup \Gamma)$ where $D(z_i,r)$ is a disc
centered at $z_i$ with small radius r and $\Gamma$ is the set lenses defined used for the second transformation;

\item $P$ has jumps:
\beq
\forall z \in \left(\mathbb{R} \cup \Gamma\right) \bigcap \left({\displaystyle \bigcup_{i=1}^{2k}} D(z_i,r)\right) \, , \; P_+(z) = P_-(z) j_S(z);
\eeq

\item When $n \to \infty$, the boundary conditions for $P$ are:
\beq
\forall i = 1, \dots, 2k \, , \; \forall z \in \partial D(z_i,r) \, , \; P(z) = \left({\bf I}+ O\left({1 \over N}\right) \right) M(z) \;\; \hbox{uniformly}.
\eeq

\end{itemize}

Let us now consider a particular branch point $z_{2i}$\footnote{We could do exactly the same for a branch point $z_{2i-1}$ with
the same function $\beta_i$.}.
When $z \to z_{2i}$, lemma \ref{lemedge} states that:
\beq
\begin{array}{c}
\l_0(z) = p_{2i}(z-z_{2i}) + {2 \rho_{2i} \over 3} (z- z_{2i})^{3\over 2} + O\left((z-z_{2i})^2\right) \cr
\l_i(z) = p_{2i}(z-z_{2i}) - {2 \rho_{2i} \over 3} (z- z_{2i})^{3\over 2} + O\left((z-z_{2i})^2\right) \cr
\end{array}
\eeq
i.e.
\beq
\l_0(z)-\l_i(z) = {4 \rho_{2i} \over 3} (z -z_{2i})^{3\over 2} + O\left((z-z_{2i})^{5 \over 2}\right)
\eeq
which means that the function $\beta_i(z) = \left[{3 \over 4}(\l_0(z)-\l_i(z))\right]^{2 \over 3}$ is analytic in a neighborhood of $z_{2i}$ with a
strictly positive derivative $\beta_i'(z_{2i}) = \rho_{2i}^{2\over 3}>0$.

Thus $\beta_i$ is a conformal map from $D(z_{2i},r)$ to a neighborhood
of the origin for $r$ small enough.

\begin{figure}
\begin{center}
\includegraphics[width=5cm]{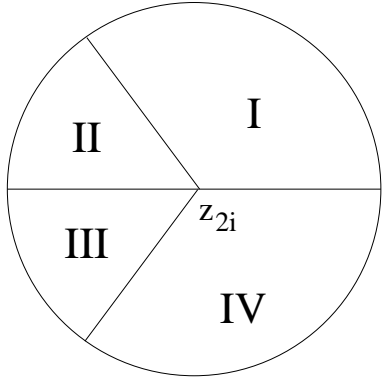}\\
\end{center}
  \caption{Decomposition of the disc $D(z_{2i},r)$ into four sectors.}\label{disc}
\end{figure}

Let us now cut the image of $D(z_{2i},r)$ by a a special choice of $\Gamma$ representing the cuts of the lens based in $z_{2i}$:
\beq
\beta_i( \Gamma \bigcap D(z_{2i},r)) = \left\{z |\hbox{arg}(z) = \pm {2 \pi \over 3} \right\} .
\eeq
It gives a partition of the disc $D(z_{2i},r)$ into four regions labeled I, II, III and  IV (see figure \ref{disc}) following:
\begin{itemize}
\item $0<\hbox{arg}(\beta(z))<{2 \pi \over 3}$ in region I;
\item ${2 \pi \over 3}<\hbox{arg}(\beta(z))<\pi$ in region II;
\item $- \pi<\hbox{arg}(\beta(z))<-{2 \pi \over 3}$ in region III;
\item $-{2 \pi \over 3}<\hbox{arg}(\beta(z))<0$ in region IV.
\end{itemize}

We can now transform the RH for $P$ to a model one by the transformation:
\beq
\widetilde{P}(z) := \left\{ \begin{array}{l}
P(z) T_{\widetilde{P} P}(z) \; \; \hbox{in regions I and IV.} \cr
P(z) \; \; \hbox{in regions II and III.} \cr
\end{array}
\right.
\eeq
where one has defined:
\beq
T_{\widetilde{P} P}(z) = \left[ \delta_{\alpha \beta} - \delta_{\alpha_i} \sum_{j \neq 0,i} \delta_{\beta j} e^{N(\l_j-\l_i)}\right]_{\alpha,\beta = 0}^{k}.
\eeq
Thus, $\widetilde{P}$ is solution of the same RH problem as $P$ except that its jumps are now given by
\beq
\forall z \in \left(\mathbb{R} \cup \Gamma\right) \bigcap \left({\displaystyle \bigcup_{i=1}^{2k}} D(z_i,r)\right) \, , \; P_+(z) = P_-(z) j_{\widetilde{P}}(z)
\eeq
where $j_{\widetilde{P}}(z)$ is given by:
\begin{itemize}
\item for $z$ on $[z_{2i}-r,z_{2i}[$:
\beq
j_{\widetilde{P}} = \left[ (1 - \delta_{\alpha 0}- \delta_{\alpha i}) \delta_{\alpha \beta} + \delta_{\alpha 0} \delta_{\beta i}-  \delta_{\alpha i} \delta_{\beta 0}\right]_{\alpha,\beta = 0}^{k};
\eeq

\item for $z$ on the upper and lower sides of the lens in $D(z_{2i},r)$:
\beq
j_{\widetilde{P}} = \left[ \delta_{\alpha \beta} +  \delta_{\alpha i} \delta_{\beta 0} e^{N(\l_0-\l_i)} \right]_{\alpha,\beta = 0}^{k};
\eeq

\item for $z$ on $]z_{2i},z_{2i}+r]$:
\beq
j_{\widetilde{P}} = \left[ \delta_{\alpha \beta} +  \delta_{\alpha 0} \delta_{\beta i} e^{N(\l_0-\l_i)} \right]_{\alpha,\beta = 0}^{k}.
\eeq
\end{itemize}

One sees that these jumps are equal to the identity except for a $2\times 2$ block corresponding to the $(0,0)$, $(0,i)$, $(i,0)$ and $(0,0)$
coefficients. The RH problem reduces then to a model problem of size $2 \times 2$ which can be solved explicitly as follows
(see \cite{Kuijlaars1} for example):
\beq\label{formPtilde}\encadremath{
\widetilde{P}(z) = E_N(z) \Phi\left(N^{2\over 3} \beta_i(z)\right) D(z)}
\eeq
where $E_N$ ensures the boundary conditions
\beq
E_N = \sqrt{\pi} M \widetilde{M} \widehat{M},
\eeq
with
\beq
\widetilde{M} = \left[ (1 - \delta_{\alpha i}(1+i)) \delta_{\alpha \beta}  -  \delta_{\alpha 0} \delta_{\beta i} - i \delta_{\alpha i} \delta_{\beta 0}  \right]_{\alpha,\beta = 0}^{k},
\eeq
and
\beq
\widehat{M} = \left[ (1 - \delta_{\alpha 0}(1-n^{1 \over 6} \beta^{1 \over 4})- \delta_{\alpha i}(1-n^{-{1 \over 6}} \beta^{-{1 \over 4}})) \delta_{\alpha \beta}  \right]_{\alpha,\beta = 0}^{k},
\eeq
$\Phi(z)$ is defined with the use of the Airy function $Ai(z)$:
\beq
 \left\{\begin{array}{l}
\left[ (1 - \delta_{\alpha 0}(1 - y_0(z)) - \delta_{\alpha i}(1 + y_2'(z)))) \delta_{\alpha \beta} + \delta_{\alpha_i} \delta_{\beta 0} y_0'(z) - \delta_{\alpha_i} \delta_{\beta 0} y_2(z)  \right]_{\alpha,\beta = 0}^{k}
\; \hbox{for} \; 0<\hbox{arg}<{2 \pi \over 3} \cr
\left[ (1 - \delta_{\alpha 0}(1 + y_1(z)) - \delta_{\alpha i}(1 + y_2'(z)))) \delta_{\alpha \beta} - \delta_{\alpha_i} \delta_{\beta 0} y_1'(z) - \delta_{\alpha_i} \delta_{\beta 0} y_2(z)  \right]_{\alpha,\beta = 0}^{k}
\; \hbox{for} \; {2 \pi \over 3}<\hbox{arg}<\pi \cr
\left[ (1 - \delta_{\alpha 0}(1 + y_2(z)) - \delta_{\alpha i}(1 - y_1'(z)))) \delta_{\alpha \beta} - \delta_{\alpha_i} \delta_{\beta 0} y_2'(z) + \delta_{\alpha_i} \delta_{\beta 0} y_1(z)  \right]_{\alpha,\beta = 0}^{k}
\; \hbox{for} \; -\pi<\hbox{arg}<-{2 \pi \over 3} \cr
\left[ (1 - \delta_{\alpha 0}(1 - y_0(z)) - \delta_{\alpha i}(1 - y_1'(z)))) \delta_{\alpha \beta} + \delta_{\alpha_i} \delta_{\beta 0} y_0'(z) + \delta_{\alpha_i} \delta_{\beta 0} y_1(z)  \right]_{\alpha,\beta = 0}^{k}
\; \hbox{for} \; -{2 \pi \over 3}<\hbox{arg}<0 \cr
\end{array}
\right.
\eeq
where
\beq\label{defyairy}
y_\alpha = e^{2 \alpha i \pi \over 3} Ai(e^{2 \alpha i \pi \over 3}z)
\eeq for $\alpha = 0,1,2$ and $D$ is the diagonal matrix
\beq
D(z) = \left[ (1 - \delta_{\alpha 0}(1-e^{{N \over 2}(\l_0(z)-\l_i(z))}) - \delta_{\alpha i}(1-e^{-{N \over 2}(\l_0(z)-\l_i(z))})) \delta_{\alpha \beta}  \right]_{\alpha,\beta = 0}^{k}.
\eeq

\subsubsection{Final transformation}
We now have a parametrix away from the branch points ($M(z)$) and in their neighborhood ($P(z)$). It remains to put them together
in a unique function defined on the whole complex plane: this is the last transformation.

Define the matrix
\beq\label{transRS}
\begin{array}{l}
R(z) = S(z) M(z)^{-1} \;\; \hbox{for $z$ outside the disks $D(z_{i},r)$, $i=1, \dots, 2k$} \cr
R(z) = S(z) P(z)^{-1} \;\; \hbox{for $z$ inside the disks $D(z_{i},r)$, $i=1, \dots, 2k$}. \cr
\end{array}
\eeq

One can easily check that:
\beq
R(z) = {\bf I} + O\left( {1 \over N(|z|+1)}\right)
\eeq
as $N \to \infty$ uniformly for $z \in \mathbb{C} \backslash \left({\displaystyle \bigcup_{i=1}^{2k}} \partial D(z_{i},r)\right)$
and
\beq
R'(z) = O\left( {1 \over N(|z|+1)}\right).
\eeq

It follows the important relation:
\beq\label{RRi}\encadremath{
R^{-1}(x_2) R(x_1) = I + O\left({x_1-x_2 \over N}\right)}
\eeq
needed to study the behavior of the Kernel in the different regimes.

\subsection{Mean density of particles}

With the use of the kernel $K_N$, one can state more precisely the role played by the spectral curve.
It follows from:
\bt
The limiting mean density of eigenvalues of the random matrix
\beq
\rho(x) := \lim_{N \to \infty} {K_N(x,x) \over N}
\eeq
exists, is supported by the real cuts $\left\{[z_{2i-1},z_{2i}]\right\}_{i=1}^k$ and is expressed in terms of the solution $\xi_0$ of the algebraic equation
\beq
{\cal{E}}(x,\xi) = 0
\eeq
by
\beq
\rho(x) = {1 \over \pi} \left|Im \xi_0(x) \right|.
\eeq
It is analytic on the cuts and vanishes like a square root at the edges $z_i$.
\et

\proof{
Let us consider $x\in ]z_{2i-1},z_{2i}[$ for any $i = 1, \dots , k$, assuming that $x$ does not belong to any $D(z_j,r)$.
Thus, \eq{transRS} and \eq{RRi} imply that
\beq
S_+^{-1}(y) S_+(x) = {\bf I} + O(x-y)
\eeq
uniformly in $N$ when $y \to x$. The expression of the kernel in terms of $S(x)$, \eq{kernelS}, gives:
\bea\label{limK1}
K_N(x,y) &=&
{ e^{N\left[h(y)-h(x)\right]}\over 2 i \pi (x-y)} \Omega_{1,S,i}(y) \left[{\bf I} + O(x-y)\right] \Omega_{2,S,i}^t(x) \cr
&=& e^{N\left[h(y)-h(x)\right]} \left[ {\sin\left(N \Im\left( \l_{0+}(x) - \l_{0+}(y)\right)\right) \over \pi (x-y) } + O(1) \right]\cr
\eea
uniformly in $N$. Remember that, by definition,
\beq
\rho(x) = {1 \over \pi} {d \Im \l_{0+}(x)\over dx} .
\eeq
Thus, we get
\beq
K_N(x,x) = N \rho(x) + O(1),
\eeq
proving the first part of the theorem. For $x$ outside of the cuts (or at a branch point), we easily see that
\beq
\lim_{N\to \infty} {K_N(x,x) \over N} =0
\eeq
because of the ordering of the $\l$ functions (see th.\ref{thneigh}).
}

\subsection{Behavior in the Bulk}

Let us first study the asymptotic of the kernel inside the support of eigenvalues. We recover the usual sine-kernel:

\bt
For any $x \in {\displaystyle \bigcup_{i=1}^k} ]z_{2i-1},z_{2i}[$ and any $(u,v) \in \mathbb{R}^2$, one has
\beq\encadremath{
\lim_{N \to \infty} {1 \over N \rho(x)} \widehat{K}_N\left(x+{u \over N \rho(x)},x+{v \over N \rho(x)}\right) = {\sin \pi (u-v) \over \pi (u-v)}
}\eeq
where we use the rescaled kernel:
\beq
\widehat{K}_N(x,y) = e^{N(h(x)-h(y))} K_N(x,y).
\eeq

\et

\proof{
Consider $x\in ]z_{2i-1},z_{2i}[$ for any $i = 1, \dots , k$ and define
\beq
X = x + {u \over N \rho(x)}
\qquad \hbox{and} \qquad
Y = x + {v \over N \rho(x)}.
\eeq
Then, there exists $t = x + O\left({1 \over N}\right)$ such that
\beq
N \Im(\l_{0+}(X)-\l_{0+}(Y)) = {\pi (u-v) \rho(t) \over \rho(x)}.
\eeq
From \eq{limK1}, it follows
\bea
{\widehat{K}_N(X,Y) \over N \rho(x)} &=& {\sin\left(N \Im\left( \l_{0+}(X) - \l_{0+}(Y)\right)\right) \over \pi (u-v) } + O \left({1 \over N}\right) \cr
&=& {\sin (\pi(u-v)) \over \pi (u-v)} + O\left({1 \over N}\right).\cr
\eea

}

\subsection{Behavior at the edge}

Let us now study the behavior of the kernel at the edge of the spectrum, i.e. near any branch point of the spectral curve.

\bt
For any $i = 1, \dots, 2k$ and every $(u,v) \in \mathbb{R}^2$, the rescaled kernel satisfies:
\beq
\lim_{N \to \infty} {1 \over \left(\rho_i N\right)^{2 \over 3}} \widehat{K}_N\left( z_i + (-1)^i {u \over \left(\rho_i N\right)^{2 \over 3}},
z_i + (-1)^i {v \over \left(\rho_i N\right)^{2 \over 3}}\right) = {Ai(u) Ai'(v) - Ai'(u) Ai(v) \over u-v},
\eeq
where $\rho_i$ is a constant characterizing the behavior of $\rho$ near the branch point $z_i$:
\beq
\rho(x) = {\rho_i \over \pi} |x-z_i|^{1 \over 2} \left( 1 + o(1)\right)
\eeq
when $x \to z_i$.

\et

\proof{
Let us consider an arbitrary branch point\footnote{We consider here a branch point on the right side of a cut but we could do
exactly the same derivation for the other end point of the cut $z_{2i-1}$.} $z_{2i}$ and define
\beq
X:= z_{2i} + {u \over \left(\rho_{2i} N\right)^{2 \over 3}}
\;\;\; \hbox{and} \;\;\;
Y:=z_{2i} + {v \over \left(\rho_{2i} N\right)^{2 \over 3}}
\eeq
for any $(u,v) \in ]0,\infty[^2$ so that, for $N$ large enough, $X$ and $Y$ lie inside the circle $D(z_{2i},r)$ on the cut.
The rescaled kernel can thus be written:
\beq
\widehat{K}_N(X,Y) = { 1 \over 2 i \pi (X-Y)} \Omega_{1,S,i}(Y) S_+^{-1}(Y) S_+(X) \Omega_{2,S,i}^t(X)
\eeq
with the notations of \eq{kernelS} and
\beq
S_+(X) = R(X) P_+(X) = R(X) \widetilde{P}_+(X)
\eeq
since we are in the region II. Using the form \ref{formPtilde}of $\widetilde{P}$, one gets:
\beq
S_+(X) = R(X) E_N(X) \Phi_+\left(N^{2 \over3} \beta_i(X)\right) D_{i,+}(X)
\eeq
where
\beq
D_{i,+}(X) = \hbox{diag}\left(  e^{N i \Im\left[\l_{0+}(X)\right]} , \overbrace{1 , \dots , 1}^{i-1} , e^{-N i \Im\left[\l_{0+}(X)\right]} ,\overbrace{1 , \dots ,1}^{k-i}\right).
\eeq
It follows
\beq
{1 \over (\rho_{2i} N)^{2 \over 3}} \widehat{K}_N(X,Y) = {1 \over 2 i \pi (u-v)} \widetilde{D}_{i} \Phi_+^{-1}\left(N^{2 \over3} \beta_i(Y)\right)
E_N^{-1}(Y) R^{-1}(Y) R(X) E_N(X) \Phi_+\left(N^{2 \over3} \beta_i(X)\right) \widehat{D}_i^t
\eeq
where we use the vector of size $k+1$:
\beq
\widetilde{D}_i:= \left[ -1 \; \overbrace{0 \; \dots \; 0}^{i-1} \; 1 \; \overbrace{0 \; \dots \; 0}^{k-i} \right]
\eeq
and
\beq
\widehat{D}_i:= \left[ 1 \; \overbrace{0 \; \dots \; 0}^{i-1} \; 1 \; \overbrace{0 \; \dots \; 0}^{k-i} \right].
\eeq
Remember that $\beta_i(z_{2i}) = 0$ and $\beta_i'(z_{2i})= \rho_{2i}^{2 \over 3}$. Thus
\beq
\Phi_+\left(N^{2 \over 3} \beta_i(X)\right) \to \Phi_+(u) \;\;\; \hbox{when} \;\;\; N \to \infty
\eeq
and
\beq
\lim_{N \to \infty} \Phi_+\left(N^{2 \over3} \beta_i(X)\right) \widehat{D}_i^t =
\left[ y_0(u) \; \overbrace{0 \; \dots \; 0}^{i-1} \; y_0'(u) \; \overbrace{0 \; \dots \; 0}^{k-i} \right]^t
\eeq
as well as
\beq
\lim_{N \to \infty} \widetilde{D}_{i} \Phi_+^{-1}\left(N^{2 \over3} \beta_i(Y)\right) =
- 2 i \pi \left[ -y_0'(u) \; \overbrace{0 \; \dots \; 0}^{i-1} \; y_0(u) \; \overbrace{0 \; \dots \; 0}^{k-i} \right]
\eeq
with the function $y_0$ defined by \eq{defyairy} in terms of the Airy function.

Since one knows the behavior of $E_N$, $E^{-1}_N$ and $R$ when $N \to \infty$:
\bea
E_N(X) &=& O\left(N^{1 \over 6}\right) \cr
E_N^{-1}(Y) &=& O\left(N^{1 \over 6}\right) \cr
E_N^{-1}(Y) E_N(X) &=& I + O\left(N^{-{1 \over 3}}\right) \cr
R^{-1}(Y) R(X) &=& I + O\left(N^{-{5 \over 3}}\right)\cr
\eea
we finally prove that
\beq
\lim_{N \to \infty} {1 \over (\rho_{2i} N)^{2 \over 3}} \widehat{K}_N(X,Y) = {y_0(u) y_0'(v) - y_0'(u) y_0(v) \over u-v}
\eeq
which gives the theorem for $u,v>0$.

For the other signs of $u$ and $v$, we can derive the result in the same way by using the appropriate expression of the kernel
as well as the following definition of the $h$ function outside of the cut.
}

\section{Intermediate state}\label{secinter}

We consider here one intermediate configuration where two or more segment have already merged.

It corresponds to having at least two non-real complex conjugate branch points.

\subsection{Study of the spectral curve}

We consider an intermediate time in the evolution of the Brownian motions. The spectral curve remains the same:
\beq
{\cal{E}}(x,y) = 0
\eeq
but the value of the coefficients makes its structure different through the position of the branch points.

Let us consider the following case (which is generic). The $2k$ branch points, solution of the equation in $y$
\beq
1 = \sum_{i=1}^k {n_i \over N\left(y-a_i(t)\right)^2}
\eeq
are such that

\begin{itemize}
\item there are $2l$ real branch points $z_i$ with $i=1, \dots, 2l$ and $0<l<k$;

\item for $i=1, \dots, l$, there are $0 \leq b_i$ pairs of distinct non-real complex conjugated branch points $\left(w_i^{(m)},\overline{w}_i^{(m)}\right)_{m=1}^{b_i}$
linked by a cut $\Gamma_i^{(m)}$ crossing the real axis in a unique point $r_i^{(m)}$ satisfying:
\beq
\forall i \, , \;  z_{2i-1}<r_i^{(1)}<r_i^{(2)}<\dots < r_i^{(b_i)}<z_{2i}.
\eeq
\end{itemize}

\begin{figure}
\hspace{1cm}  \includegraphics[width=14cm]{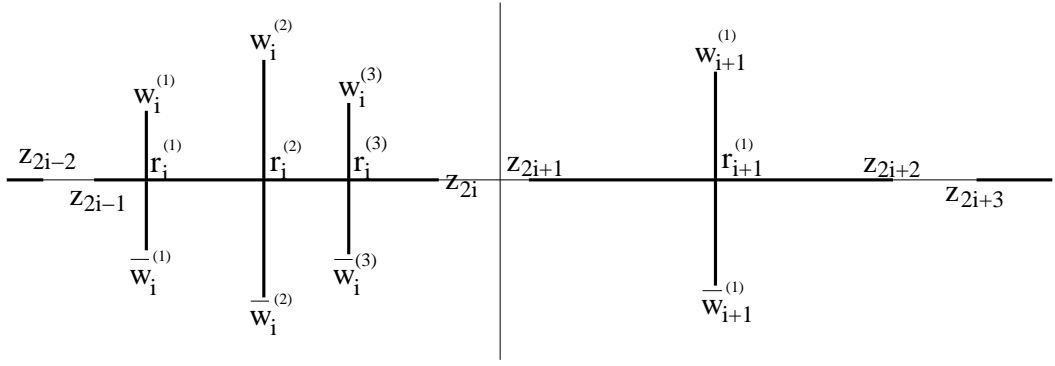}\\
  \caption{Cuts of the $\xi$ functions in the $x$-plane}\label{interspecx}
\end{figure}

\br
One obviously has the relation:
\beq
k = l + \sum_{i=1}^l b_i.
\eeq
and we use a double index notation to sort the filling fraction:
\beq
n_i^{(m)}:= n_{\sum_{j=1}^{i-1}b_j +m+i}
\eeq
and the eigenvalues:
\beq
a_i^{(m)}:= a_{\sum_{j=1}^{i-1}b_j +m+i}.
\eeq

\er

Let us now study the properties of the spectral curve as we did in the preceding section. We are particularly interested in its
sheeted structure.

A generic point of the complex plane (i.e. not a branch point) has $k+1$ pre-images on the spectral curve. Let us note these
pre-images by $\xi_0(x)$ and $\xi_i^{(m)}(x)$ with $i=1, \dots, l$ and $m=0, \dots, b_i$.
These different pre-images are characterized by their asymptotic behaviors when $x\to \infty$:
\beq
\begin{array}{c}
\xi_0(x) = x - {1 \over x}+ O\left({1 \over x^2}\right), \cr
\xi_i^{(m)} = a_{i}^{(m)} + {n_{i}^{(m)} \over N} {1 \over x} +  O\left({1 \over x^2}\right). \cr
\end{array}
\eeq
From these asymptotic behaviors, one can define the $\xi$-functions by analytical continuation as follows\footnote{For the sake
of brevity, I did not include any proof of this sheet structure in the present manuscript. Nevertheless, the simplicity of the
spectral curve implies a simple, but long and uninteresting, proof of these facts. One can, for example, follow the time evolution
of the sheet structure from $t=1$ to $t=0$ through the rational parametrization of the algebraic curve. As such a proof does not
involve any new technical element, it is left to the reader as an exercise.}:
\begin{itemize}
\item $\xi_0(x)$ can be analytically continued to $\mathbb{C} \backslash {\displaystyle \bigcup_{i=1}^l} [z_{2i-1},z_{2i}]$;

\item for $i=1, \dots, l$ and $m=0, \dots, b_i$, $\xi_i^{(m)}(x)$ can be analytically continued to
$\mathbb{C} \backslash \left(\Gamma_i^{(m)} \cup [r_i^{(m)},r_{i}^{m+1}] \cup \Gamma_i^{(m+1)}\right)$
where we use the notation:
\beq
\left\{\begin{array}{l}
w_i^{(0)} = \overline{w}_i^{(0)} = r_i^{(0)} := z_{2i-1}, \cr
w_i^{(b_i+1)} = \overline{w}_i^{(b_i+1)} = r_i^{(b_i+1)} := z_{2i}.\cr
\end{array}
\right.
\eeq

\end{itemize}

\begin{figure}
\begin{center}
\includegraphics[width=9cm]{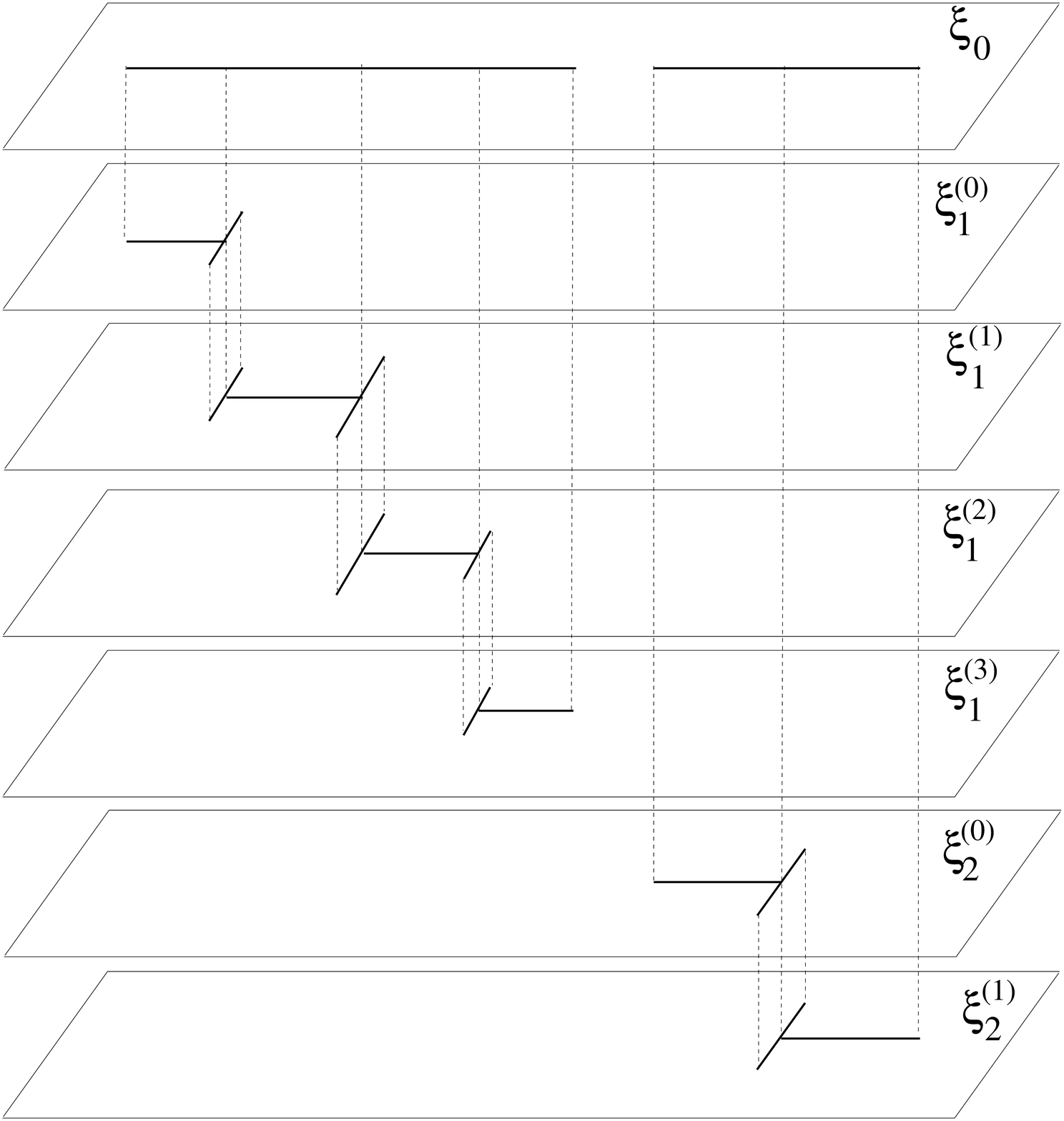}\\
\end{center}
  \caption{Sheeted structure of the spectral curve for an intermediate step. In this example, there are 4 real branch points
and 4 pairs of conjugated ones.}\label{interspec}
\end{figure}

It implies the sheet structure described in fig.\ref{interspec}\footnote{We label the sheets (except the $0$-sheet) by two
indices $(i,m)$ corresponding to the associated $\xi$ function $\xi_i^{(m)}$.}:
\begin{itemize}
\item the $0$-sheet is connected to the $(i,m)$-sheet along the real cut $[r_i^{(m)},r_i^{(m+1)}]$ for $i=1, \dots, l$ and $m=0, \dots, b_i$
with the jumps:
\beq
\xi_{0\pm}(x) = \xi_{i\mp}^{(m)}(x) \;\; \hbox{for} \;\;  x \in ]r_i^{(m)},r_i^{(m+1)}[;
\eeq

\item for $i=1, \dots, l$ and $m=1, \dots, b_i-1$, the $(i,m)$-sheet is connected to the $(i,m-1)$- and $(i,m+1)$-sheets
along the imaginary cuts $\Gamma_i^{(m)}$ and $\Gamma_i^{(m+1)}$ respectively, with the
jumps:
\beq
\begin{array}{c}
\xi_{i\pm}^{(m)}(x) = \xi_{i\mp}^{(m-1)}(x) \;\; \hbox{for} \;\; x \in \Gamma_i^{(m)}, \cr
\xi_{i\pm}^{(m)}(x) = \xi_{i\mp}^{(m+1)}(x) \;\; \hbox{for} \;\; x \in \Gamma_i^{(m+1)}. \cr
\end{array}
\eeq

\end{itemize}

We finally define the associated $\l$ functions as in the preceding section:
\bd
Let the function $\l_i^{(m)}$ be the primitives of $\xi_i^{(m)}$ defined by the conditions:
\beq\label{condinitl0inter}
\l_0(z_{2l}) = 0,
\eeq
\beq
\l_{i+}^{(0)}(z_{2i-1}) = \l_{0-}(z_{2i-1}),
\eeq
\beq
\l_{i}^{(b_i)}(z_{2i}) = \l_{0+}(z_{2i})
\eeq
and
\beq\label{condinitliinter}
\l_{i+}^{(m)}(w_i^{(m+1)}) = \l_{i-}^{(m+1)}(w_i^{(m+1)})
\eeq
for $m= 1, \dots , b_i-1$.
\ed

One can easily prove that they satisfy the following properties:
\bl
$\l_0$ is analytic in $\mathbb{C} \backslash {\displaystyle \bigcup_{i}} [z_{2i-1},z_{2i}]$ and
$\l_i^{(m)}$ is analytic in $\mathbb{C} \backslash {\displaystyle \bigcup_{j=1}^{i-1}} \left[z_{2j-1},z_{2j}\right] \cup \left[z_{2i-1},r_i^{(m)}\right] \cup \Gamma_i^{(m)}$
for $i=1,\dots, k$ and $m=1, \dots, b_i$.

The jumps and asymptotics of these $\l$-functions are given by:
\begin{itemize}
\item When $z \to \infty$:
\beq
\l_0 = {z^2\over 2} - \ln z + l_0 + O\left({1 \over z^2}\right)
\eeq
and, for $i \neq 0$:
\beq
\l_i^{(m)} = a_{i}^{(m)} z + {n_i^{(m)} \over N} \ln z  + l_i^{(m)} + O\left({1 \over z^2}\right)
\eeq
where $l_i^{(m)}$'s are some constants chosen to satisfy the constraints $\left(\ref{condinitl0inter}\right)-\left(\ref{condinitliinter}\right)$.

\item They follow the jumps:
\bea
\l_{0+}(x) - \l_{0-}(x) = - 2 i \pi \sum_{j=i+1}^l \sum_{m=1}^{b_j} {n_{j}^{(m)} \over N}, & \;\; & x \in ]z_{2i-2},z_{2i-1}[ \cr
\l_{0+}(x) - \l_{0-}(x) = - 2 i \pi, & \;\; & x \in ]-\infty,z_1[. \cr
\eea
For $i \in [1,l]$
\bea
\l_{i+}^{(m)}(x) - \l_{i-}^{(m)}(x) = 2 i \pi {n_{i}^{(m)} \over N},& \;\; & x \in ]-\infty,r_{i}^{(m)}[ \cr
\l_{0+}(x) - \l_{l-}^{(b_l)}(x) = 0 \, , \; \l_{0-}(x) - \l_{l+}^{(b_l)}(x) = 0, & \;\; & x \in ]z_{2l-1},z_{2l}[ \cr
\eea
For $i \in [1,l-1]$
\bea
\l_{0+}(x) - \l_{i-}^{(m)}(x) = 0,  & \;\; & x \in [r_{i}^{(m)},r_{i}^{(m+1)}] \cr
\eea
and
\bea
\l_{i\pm}^{(m)}(x) = \l_{i\mp}^{(m+1)}(x), & \;\; & x \in \Gamma_i^{(m)+}
\eea
where $\Gamma_i^{(m)+}$ denotes the part of $\Gamma_i^{(m)}$ lying inside the upper half plane.

\end{itemize}

\el

\subsubsection{Study of the sign of $\Re\left(\l_i^{(m)}-\l_j^{(h)}\right)$}

It is very important to know the sign of $\Re\left(\l_i^{(m)}-\l_j^{(h)}\right)$ for $(i,m)\neq (j,h)$ in order to study the asymptotic behavior
of the jump matrices in the Riemann-Hilbert problem we want to solve. For this purpose, it is convenient to draw the curves
where $\Re\left(\l_i^{(m)}\right)=\Re\left(\l_j^{(h)}\right)$. Let us build them step by step.

\begin{figure}
\begin{center}
  \includegraphics[width=6.5cm]{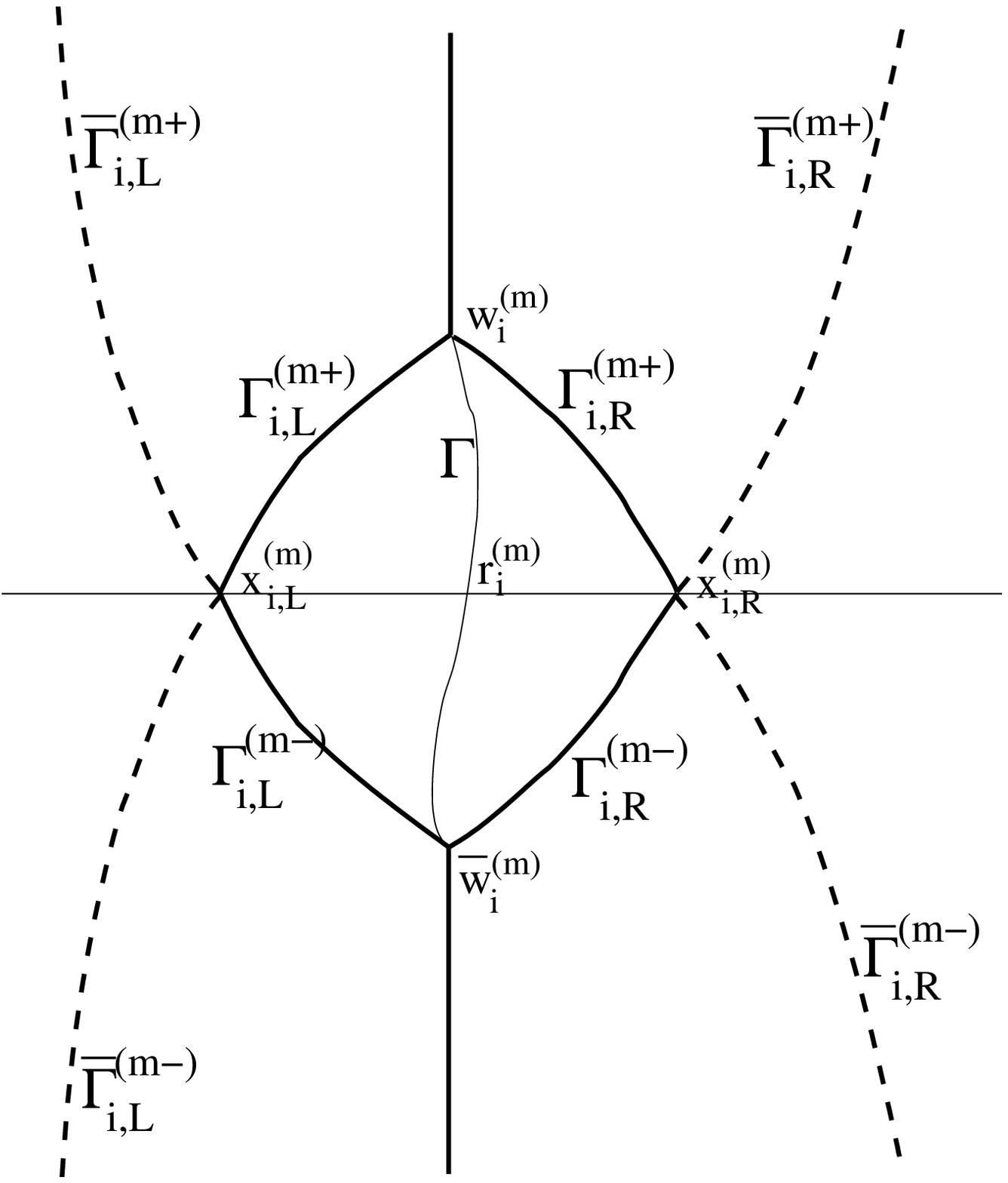}\\
\end{center}
  \caption{Zoom near the vertical cut $\Gamma_i^{(m)}$. The thick solid lines are the curves where $\Re \left(\l_i^{(m)}-\l_i^{(m-1)} \right)=0$.
On the dashed lines $\overline{\Gamma}_{iR}^{m\pm}$ (resp. $\overline{\Gamma}_{iL}^{m\pm}$), one has $\Re \left(\l_0-\l_i^{(m-1)} \right)=0$
(resp. $\Re \left(\l_i^{(m)}-\l_0 \right)=0$).  }\label{figgammaim}
\end{figure}

First of all, from the behavior of the $\l$-functions near the branch points, one can state the following results:
\begin{itemize}
\item From any branch point $z_{2i-1}$ go three curves at equal angles ${2 \pi \over 3}$ where $\Re\left(\l_0\right) = \Re\left(\l_i^{(0)}\right)$,
one of them being $\left[z_{2i-1},r_i^{(1)}\right]$.

\item From any branch point $z_{2i}$ go three curves at equal angles ${2 \pi \over 3}$ where $\Re\left(\l_0\right) = \Re\left(\l_i^{(b_i)}\right)$
one of them being $\left[r_i^{(b_i)},z_{2i}\right]$.

\item From any branch point $w_{i}^{(m)}$ (resp. $\overline{w}_{i}^{(m)}$) go three curves at equal angles ${2 \pi \over 3}$ where
 $\Re\left(\l_i^{(m-1)}\right) = \Re\left(\l_i^{(m)}\right)$, one of them being $\left[w_i^{(m)},i \infty \right[$
(resp. $\left[\overline{w}_i^{(m)},-i \infty \right[$). Let us denote $\Gamma_{i,R}^{(m+)}$ and $\Gamma_{i,L}^{(m+)}$
 (resp. $\Gamma_{i,R}^{(m-)}$ and $\Gamma_{i,L}^{(m-)}$) the two other curve going from $w_{i}^{(m)}$ (resp. $\overline{w}_{i}^{(m)}$),
the first one lying to the right hand side of the curve $\Gamma_i^{(m)}$ (see fig. \ref{figgammaim}). The curves $\Gamma_{i,R}^{(m+)}$
and $\Gamma_{i,R}^{(m-)}$ (resp. $\Gamma_{i,L}^{(m+)}$
and $\Gamma_{i,L}^{(m-)}$) intersect in the real line in $x_{i,R}^{(m)}$ (resp. $x_{i,L}^{(m)}$). On the analytical continuation
$\overline{\Gamma}_{i,R}^{(m\mp)}$ (resp. $\overline{\Gamma}_{i,L}^{(m\mp)}$) of $\Gamma_{i,R}^{(m\pm)}$ (resp. $\Gamma_{i,L}^{(m\pm)}$)
beyond $x_{i,R}^{(m)}$ (resp. $x_{i,L}^{(m)}$), one has $\Re\left(\l_0 - \l_i^{(m-1)}\right) = 0$ (resp. $\Re\left(\l_0 - \l_i^{(m)}\right) = 0$)
because of the cut on the real line.

\end{itemize}

We have investigated the curves $\Re\left(\l_i^{(m)}-\l_j^{(h)}\right)$ near the branch points, let us now look at their
behavior away from them. Many different configuration can occur, depending on the relative positions of the $x_{i,R}^{(m)}$'s,
the $x_{i,L}^{(m)}$'s, the $r_i^{(m)}$'s and
the real branch points $z_i$. In order to limit the size of this paper, we restrict our study to the cases where
$r_i^{(m)}<x_{i,R}^{(m)}<r_i^{(m+1)}$ and $r_i^{(m)}<x_{i,L}^{(m+1)}<r_i^{(m+1)}$. Nevertheless, the RH problem corresponding
to the other possible configurations can be solved with the same method and the same transformations even if we do not give a proof here
(the study if the signs of $\Re \left(\l_i^{(m)}-\l_j^{(h)}\right)$ is a little bit more subtle but does not present any particular
difficulty).

Note also that the curves $\Re\left(\l_i^{(m-1)}\right) = \Re\left(\l_i^{(m)}\right)$ and $\Re\left(\l_i^{(m+1)}\right) = \Re\left(\l_i^{(m)}\right)$
cross in two points symmetric with respect to the real axis. On the line $D_i^{\left(m+{1\over 2}\right)}$ joining these two points,
one has $\Re\left(\l_i^{(m-1)}\right) = \Re\left(\l_i^{(m+1)}\right)$.

\begin{figure}
\begin{center}
  \includegraphics[width=12cm]{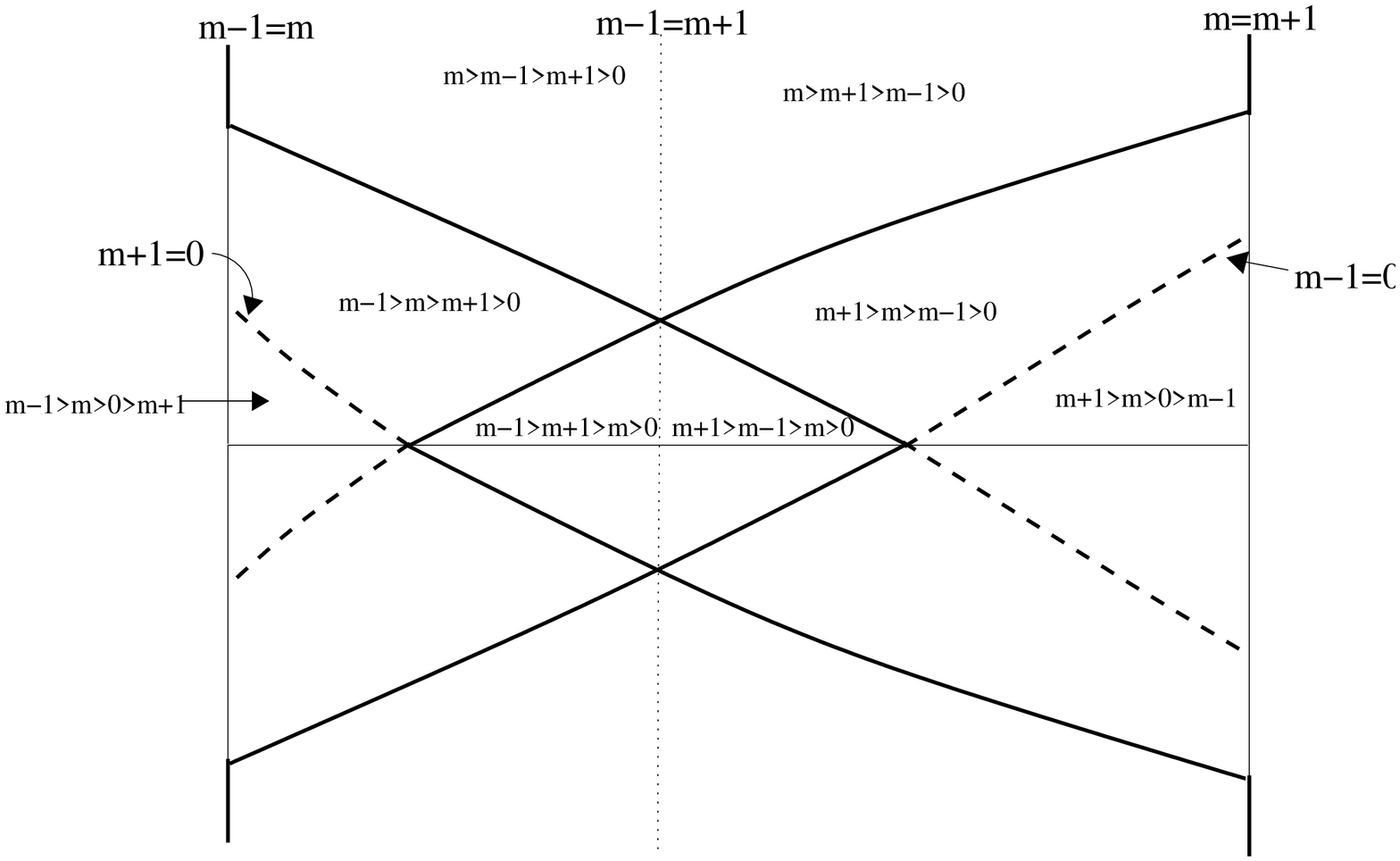}\\
\end{center}
  \caption{Study of the sign of $\Re\left(\l_j^{(h)}-\l_{j'}^{(h')}\right)$ for the second case. For readability,
$m>m+1>m-1>0$ stands for $\Re\left(\l_i^{m}(x)\right)>\Re\left(\l_i^{m+1}(x)\right)>\Re\left(\l_i^{m-1}(x)\right)>\Re\left(\l_0^{0}(x)\right)$.}\label{sign2}
\end{figure}

With this assumption, between $r_i^{(m)}$ and $r_i^{(m+1)}$, there are two different possible configurations:
\begin{itemize}
\item In the first case:
\beq
r_i^{(m)}<x_{i,R}^{(m)}<x_{i,L}^{(m+1)}<r_i^{(m+1)}.
\eeq
The respective weights of $\Re\left(\l_i^{m-1}(x)\right)$, $\Re\left(\l_i^{m}(x)\right)$, $\Re\left(\l_i^{m+1}(x)\right)$
and $\Re\left(\l_0^{(0)}(x)\right)$ for $x$ between $\Gamma_i^{(m)}$ and $\Gamma_i^{(m=1)}$ are depicted in fig.\ref{sign1}.

\begin{figure}
\begin{center}
  \includegraphics[width=9cm]{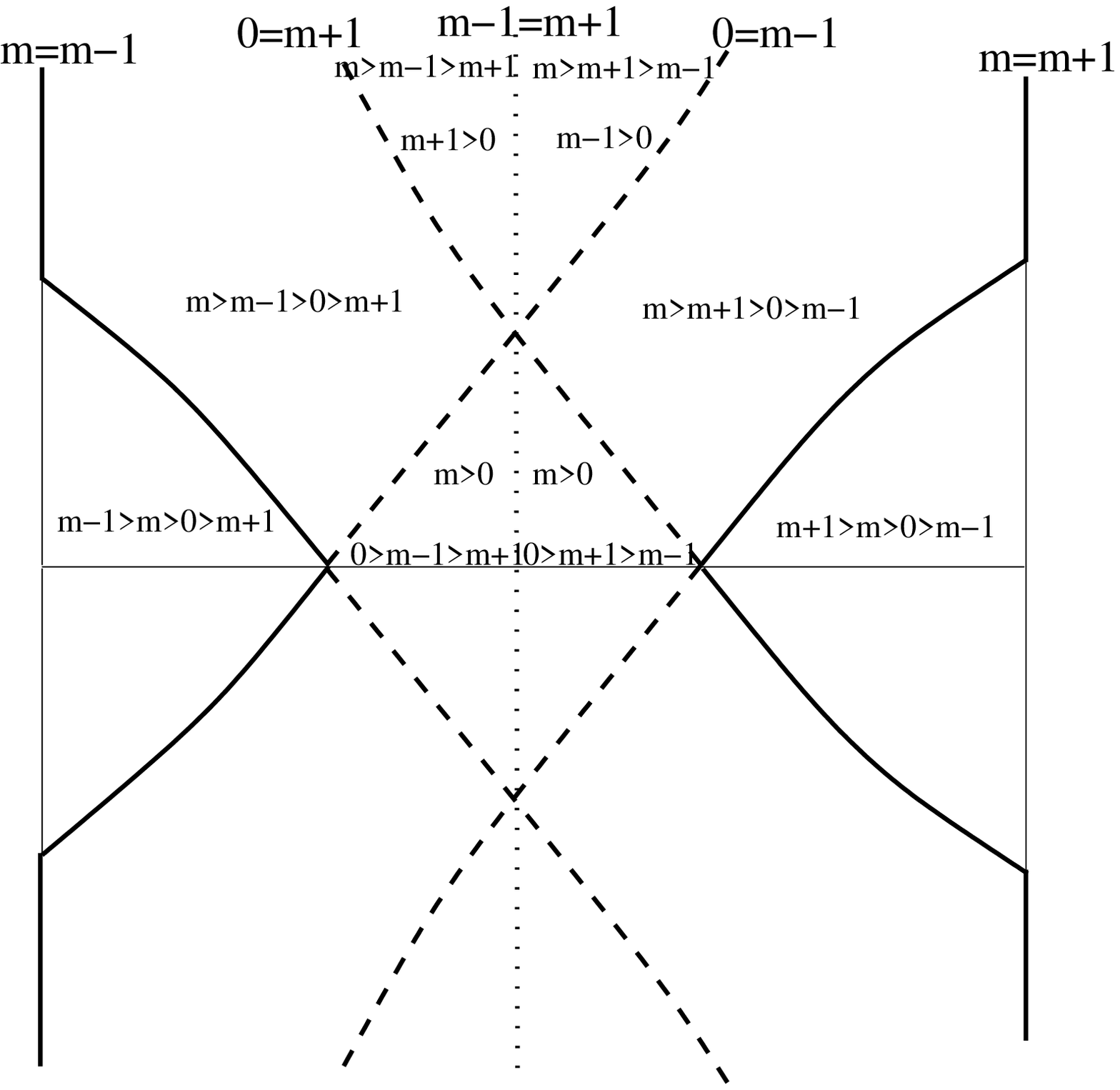}\\
\end{center}
  \caption{Study of the sign of $\Re\left(\l_j^{(h)}-\l_{j'}^{(h')}\right)$ for the first case. For readability,
$m>m+1>m-1>0$ stands for $\Re\left(\l_i^{m}(x)\right)>\Re\left(\l_i^{m+1}(x)\right)>\Re\left(\l_i^{m-1}(x)\right)>\Re\left(\l_0^{0}(x)\right)$. }\label{sign1}
\end{figure}

\item In the second case:
\beq
r_i^{(m)}<x_{i,L}^{(m+1)}<x_{i,R}^{(m)}<r_i^{(m+1)}.
\eeq
The respective weights of $\Re\left(\l_i^{m-1}(x)\right)$, $\Re\left(\l_i^{m}(x)\right)$, $\Re\left(\l_i^{m+1}(x)\right)$
and $\Re\left(\l_0^{(0)}(x)\right)$ for $x$ between $\Gamma_i^{(m)}$ and $\Gamma_i^{(m=1)}$ are depicted in fig.\ref{sign2}.

\end{itemize}

In both configuration one has the following result:
\bl
For $x$ between $\Gamma_i^{(m)}$ and $\Gamma_i^{(m+1)}$:
\beq
\l_j^{(h)}(x)>\l_0(x)
\eeq
for $(j,h)\neq(i,m-1),(i,m),(i,m+1)$.

\el

\subsection{Riemann-Hilbert analysis}
 We now solve the Riemann Hilbert problem presented in the preceding section for this particular value of the external matrix eigenvalues $a_i$ by performing a sequence of transformation leading to a simple problem when $N \to \infty$.

For convenience, we use a double index notation for the matrices in this section: considering an arbitrary $k+1 \times
k+1$ matrix $A$, we note its matrix elements as follows:
\beq
[A]_{(i,m);(i',m')}:= [A]_{\sum_{j=1}^{i-1}b_j +m+i, \sum_{j=1}^{i'-1}b_j +m'+i'}
\eeq
for $i= 1 , \dots , l$ and $m= 0, \dots, b_i$. The first line (or column) is denoted by the pair $(0,0)$.
This notation coincides with the notation used to sort the branch points and filling fraction as well as the $\xi_i^{(m)}$
and $\l_i^{(m)}$ functions: the lines are sorted in $i$ groups of $b_i$ elements corresponding to the $i$ real cuts
$[z_{2i-1},z_{2i}]$ shared by $b_i+1$ sheets.

With these notations, let us remember the RH problem to solve. We are looking for a $k+1 \times k+1$ matrix $Y(x)$
satisfying the constraints:

\begin{itemize}
\item $Y$ is analytic on $\mathbb{C} \; \backslash \; \mathbb{R}$;

\item for $x \in \mathbb{R}$, one has the jumps
\beq
Y_+(x) = Y_-(x) \left( \begin{array}{ccccc}
1 & \omega_1^{(0)}(x) & \omega_1^{(1)}(x) & \dots & \omega_l^{(b_l)}(x) \cr
0 & 1 & 0 & \dots & 0 \cr
0 & 0 & 1 & \dots & 0 \cr
\dots & \dots & \dots & \dots & \dots \cr
0 & 0 & 0 & \dots & 1 \cr
\end{array} \right)
\eeq
where  $Y_+(x)$ and $Y_-(x)$ denote respectively the limit of $Y(z)$ when
$z \to  x$ from the upper and lower half planes and $\omega_i^{(m)}(x):= e^{-N \left({x^2\over 2}-a_i^{(m)} x \right)}$.

\item when $x \to \infty$, one has the asymptotic behavior:
\beq
Y(x) = \left({\bf I} + O\left( {1 \over x}\right)\right) \left(
\begin{array}{ccccc}
z^N & 0 & 0 & \dots & 0 \cr
0 & z^{-n_1^{(0)}} & 0 & \dots & 0 \cr
0 & 0 & z^{-n_1^{(1)}} & \dots & 0 \cr
\dots & \dots & \dots & \dots & \dots \cr
0 & 0 & 0 & \dots & z^{-n_l^{(b_l)}} \cr
\end{array} \right).
\eeq
\end{itemize}

\subsubsection{First transformation}
The first transformation is the same as in the large time case when the branch points are all real.

We define:
\beq
T(x):= \Omega_{1,YT} Y(x) \Omega_{2,YT}
\eeq
for $x \in \mathbb{C} \backslash \mathbb{R}{\displaystyle \bigcup_{i=1}^{l}\bigcup_{m=1}^{b_i}} [w_i^{(m)},\overline{w}_i^{(m)}]$, with
\beq
\Omega_{1,YT}:= \hbox{diag}\left(e^{-N l_0},e^{-N l_1^{(0)}}, \dots, e^{-N l_l^{(b_l)}}\right)
\eeq
and
\beq
\Omega_{2,YT}:= \hbox{diag}\left(e^{N \left(\l_0 - {x^2 \over 2}\right)},e^{N \left(\l_1^{(0)} - {n_1^{(0)} \over N} x\right)},e^{N \left(\l_1^{(1)} - {n_1^{(1)} \over N} x\right)}, \dots, e^{N \left(\l_l^{(b_l)} - {n_l^{(b_l)} \over N} x\right)}\right).
\eeq

Then one can check that it is solution of the Riemann-Hilbert problem:
\begin{itemize}
\item $T$ is analytic on $x \in \mathbb{C} \backslash \mathbb{R}{\displaystyle \bigcup_{i=1}^{l}\bigcup_{m=1}^{b_i}} [w_i^{(m)},\overline{w}_i^{(m)}]$;

\item $T$ has the jumps
\beq
T_+(x) = T_-(x) j_T(x)
\eeq
where
\beq
j_T(x) = \left( \begin{array}{ccccc}
e^{N(\l_{0+}(x)-\l_{0-}(x))} & e^{N(\l_{1+}^{(0)}(x)-\l_{0-}(x))} & e^{N(\l_{1+}^{(1)}(x)-\l_{0-}(x))} & \dots & e^{N(\l_{l+}^{(b_l)}(x)-\l_{0-}(x))} \cr
0 & e^{N(\l_{1+}^{(0)}(x)-\l_{1-}^{(0)}(x))} & 0 & \dots & 0 \cr
0 & 0 & e^{N(\l_{1+}^{(1)}(x)-\l_{1-}^{(1)}(x))} & \dots & 0 \cr
\dots & \dots & \dots & \dots & \dots \cr
0 & 0 & 0 & \dots & e^{N(\l_{l+}^{(b_l)}(x)-\l_{l-}^{(b_l)}(x))} \cr
\end{array} \right)
\eeq
for $x \in \mathbb{R}$ and
\beq
\left[j_T(x)\right]_{(j,h);(j',h')}= \delta_{(j,h);(j',h')} \left[ 1 + \left(\delta_{(j,h);(i,m)}+\delta_{(j,h);(i,m+1)}\right)
\left(e^{N\left(\l_{j+}^{(h)}-\l_{j-}^{(h)}\right)} -1 \right) \right]
\eeq
for $x \in [w_i^{(m+1)},\overline{w}_i^{(m+1)}]$ with $i=1,\dots,l$ and $m=0,\dots b_i-1$\footnote{It simply means that the jump
matrix is the identity for $x \in [w_i^{(m+1)},\overline{w}_i^{(m+1)}]$ except for the $(i,m)$'th and $(i,m+1)$'th diagonal terms
which are equal to $e^{N\left(\l_{i+}^{(m)}-\l_{i-}^{(m)}\right)}$ and $e^{N\left(\l_{i+}^{(m+1)}-\l_{i-}^{(m+1)}\right)}$
respectively.}.

\item when $x \to \infty$, one has the asymptotic behavior:
\beq
T(x) = \left({\bf I} + O\left( {1 \over x}\right)\right).
\eeq

\end{itemize}

Note that one can factorize the jump matrix $j_T(x)$ depending on the cut on which $x$ lies:

For $x \in [r_i^{(m)},r_i^{(m+1)}]$ with $i=1,\dots, l$ and $m=0,\dots, b_i$,  all the diagonal terms of $j_T$ reduce to
1 except the first one and the term $[(i,m);(i,m)]$ whereas the term $[(0,0);(i,m)]$ (i.e. the $(i,m)$'th term
of the first line) also goes to 1\footnote{We show here the form of $j_T$ for $m\neq 0,b_i$. These two cases take exactly the same
kind of form.}:
{ \beq\label{jTri}
\left( \begin{array}{cccccccc}
e^{N(\l_{0+}-\l_{0-})} & e^{N(\l_{1+}^{(0)}-\l_{0-})} & \dots & e^{N(\l_{i+}^{(m-1)}-\l_{0-})}& 1 & e^{N(\l_{i+}^{(m+1)}-\l_{0-})}& \dots & e^{N(\l_{l+}^{(b_l)}-\l_{0-})} \cr
0 & 1 & \dots & 0 & 0 & 0 & \dots & 0 \cr
\dots &  \dots  & \dots &  \dots  &  \dots  &  \dots  & \dots &  \dots  \cr
0 &  0 & \dots & 1 & 0 & 0 & \dots & 0 \cr
0 &  0 & \dots & 0 & e^{N(\l_{i+}^{(m)}-\l_{i-}^{(m)})} & 0 & \dots & 0 \cr
0 &  0 & \dots & 0 & 0 & 1 & \dots & 0 \cr
\dots &  \dots  & \dots &  \dots  &  \dots  &  \dots  & \dots &  \dots  \cr
0 &  0 & \dots & 0 & 0 & 0 & \dots & 1 \cr
\end{array} \right).
\eeq}

For $x \in \mathbb{R} \backslash \bigcup_{i} [z_{2i-1},z_{2i}]$, all the diagonal terms of $j_T$
reduce to $1$ and it takes the form
\beq
j_T = \left( \begin{array}{ccccc}
1 & e^{N(\l_{1+}^{(0)}(x)-\l_{0-}(x))} & e^{N(\l_{1+}^{(1)}(x)-\l_{0-}(x))} & \dots & e^{N(\l_{l+}^{(b_l)}(x)-\l_{0-}(x))} \cr
0 & 1 & 0 & \dots & 0 \cr
0 & 0 & 1 & \dots & 0 \cr
\dots & \dots & \dots & \dots & \dots \cr
0 & 0 & 0 & \dots & 1 \cr
\end{array} \right).
\eeq

The kernel can be written:
\beq
K_N(x,y) = { e^{N\left({x^2 \over 4}-{y^2 \over 4}\right)}\over 2 i \pi (x-y)} \Omega_{1,T}(y) T_+^{-1}(y) T_+(x) \Omega_{2,T}^t(x)
\eeq
where
\beq
\Omega_{1,T}(x) = [0 \; e^{N \l_{1+}^{(0)}(x)} \; \dots \; e^{N \l_{l+}^{(b_l)}(x)}]
\eeq
and
\beq
\Omega_{2,T}(x) = [e^{-N \l_{0+}(x)} \; 0 \; \dots \; 0].
\eeq

\subsubsection{Second transformation, global opening of lenses}
We now proceed to the second transformation allowing to cancel the exponentially increasing coefficients of the jump matrix
$j_T$. This transformation was introduced in \cite{Kuijlaars2}, for the particular case of two imaginary branch points $i=1$ and
$b_1=1$. We generalize this procedure using intensively the study of the spectral curve presented in the preceding section.

\begin{figure}
  \begin{center}
\includegraphics[width=16cm]{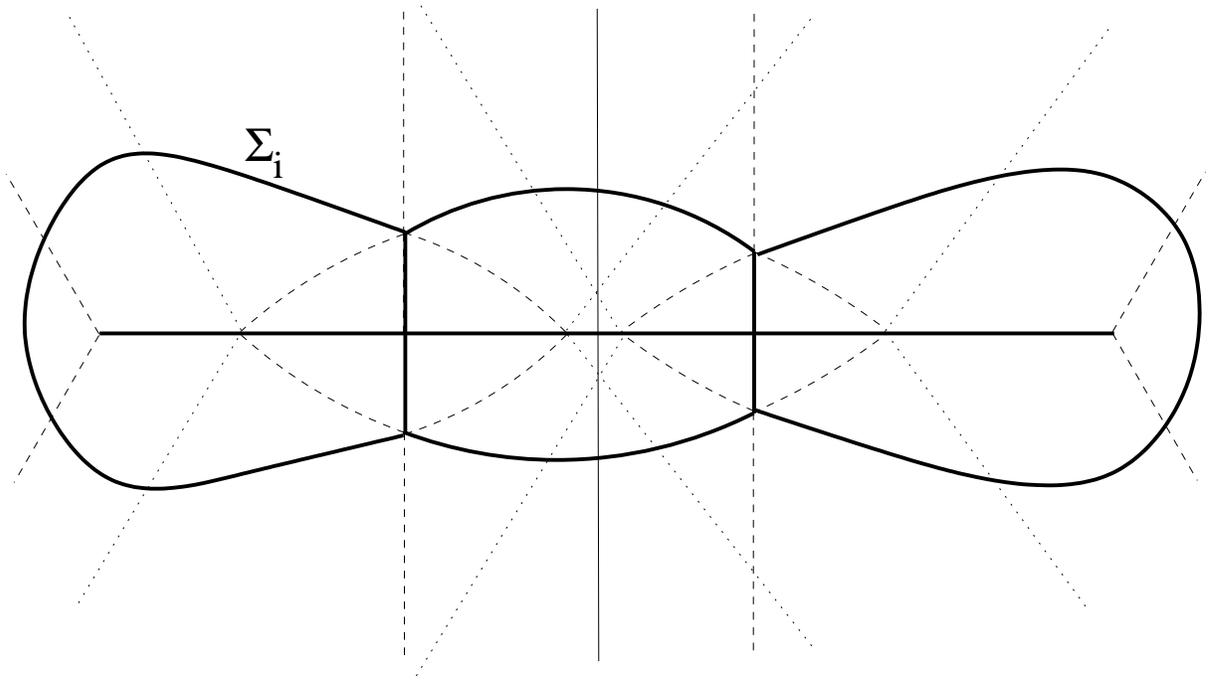}\\
\end{center}
  \caption{Example of contour $\Sigma$ used for the global opening of lenses. Here, $b_i=2$ and we are in the first case
where $x_{i,R}^{(1)}<x_{i,L}^{(2)}$.}\label{signinter}
\end{figure}

For this purpose, we define a contour $\Sigma= {\displaystyle \bigcup_{i=1}^l} \Sigma_i$ composed of $l$ connected components $\Sigma_i$
build as follows (see figures \ref{sign1}, \ref{sign2} and \ref{signinter}):
\begin{itemize}

\item $\Sigma_i$ is a closed curved encircling the cut $[z_{2i-1},z_{2i}]$ cutting the real axis in $x_{2i-1}$ and $x_{2i}$
satisfying:
\beq
x_{2i-2}<x_{2i-1}<z_{2i-1}<z_{2i}<x_{2i};
\eeq

\item $\Sigma_i$ intersects the line $r_i^{(m)}+ i \mathbb{R}$ in $w_i^{(m)}$ and $\overline{w}_i^{(m)}$ for $m=1,\dots,b_i$;

\item in a neighborhood of $w_i^{(h)}$ and $\overline{w}_i^{(m)}$, $\Sigma_i$ is the analytic continuation of
the curves $\Re\left(\l_i^{(m)}\right) = \Re\left(\l_i^{(m-1)}\right)$;

\item in the half-plane to the left (resp. to the right) of $r_i^{(m)}+ i \mathbb{R}$, $\Sigma$ lies in the region
$\Re\left(\l_i^{(m-1)}\right)>\Re\left(\l_i^{(m)}\right)$ (resp. $\Re\left(\l_i^{(m-1)}\right)<\Re\left(\l_i^{(m)}\right)$)
for any $m=1,\dots, b_i$.

\end{itemize}

Let us now define the second transformation $T \to U$.

\bd
Let $U(x)$ be the $k+1\times k+1$ matrix defined by $U(x) = T(x) J_{UT}(x)$ where $J_{UT}$ is given by
\begin{itemize}
\item For $x$ outside $\Sigma$, $J_{UT}(x) = {\bf I}$;

\item For $x$ inside $\Sigma_i$ and $\Re (x) < r_i{(1)}$:
\beq
\left[J_{UT}(x)\right]_{(j,h);(j',h')}:= \delta_{(j,h);(j',h')} - \delta_{(j,h);(i,0)} \delta_{(j',h');(i,1)} e^{N\left(\l_i^{(1)}(x)-\l_i^{(0)}(x)\right)};
\eeq

\item For $x$ inside $\Sigma_i$ between $\Gamma_i^{(m)}$ and $\Gamma_i^{(m+1)}$:
\bea
\left[J_{UT}(x)\right]_{(j,h);(j',h')} &:=& \delta_{(j,h);(j',h')} - \delta_{(j,h);(i,m)} \delta_{(j',h');(i,m-1)} e^{N\left(\l_i^{(m-1)}(x)-\l_i^{(m)}(x)\right)}\cr
&& - \delta_{(j,h);(i,m)} \delta_{(j',h');(i,m+1)} e^{N\left(\l_i^{(m+1)}(x)-\l_i^{(m)}(x)\right)};\cr
\eea

\item For $x$ inside $\Sigma_i$ and $\Re (x) > r_i{(b_i)}$:
\beq
\left[J_{UT}(x)\right]_{(j,h);(j',h')}:= \delta_{(j,h);(j',h')} - \delta_{(j,h);(i,b_i)} \delta_{(j',h');(i,b_i-1)} e^{N\left(\l_i^{(b_i-1)}(x)-\l_i^{(b_i)}(x)\right)}.
\eeq

\end{itemize}

\ed

The matrix $U(x)$ is then solution of the Riemann-Hilbert problem:
\begin{itemize}

\item $U(x)$ is analytic in $\mathbb{C} \backslash \left( \mathbb{R} \bigcup \Sigma {\displaystyle \bigcup_{i,m}} \Gamma_i^{(m)}\right)$;

\item when $x \to \infty$, one has the asymptotic behavior:
\beq
U(x) = \left({\bf I} + O\left( {1 \over x}\right)\right).
\eeq

\item One has the jumps $U_+(x) = U_-(x) j_U(x)$ with the jump matrix given by:

+ for $x \in ]r_i^{(m)},r_i^{\left(m+1\right)}[$, with $i=1,\dots,l$ and $m=0,\dots, b_i$, the jump of the matrix
$U$ is the same as jump matrix $j_T$ of \eq{jTri} except the $(i,m-1)$'th and the $(i,m+1)$'th terms of the first line which are taken equal to 0:
{\tiny \beq\label{jumpU1}
\left( \begin{array}{cccccccc}
e^{N(\l_{0+}(x)-\l_{0-}(x))} & e^{N(\l_{1+}^{(0)}(x)-\l_{0-}(x))} & \dots & 0& 1 & 0& \dots & e^{N(\l_{l+}^{(b_l)}(x)-\l_{0-}(x))} \cr
0 & 1 & \dots & 0 & 0 & 0 & \dots & 0 \cr
\dots &  \dots  & \dots &  \dots  &  \dots  &  \dots  & \dots &  \dots  \cr
0 &  0 & \dots & 1 & 0 & 0 & \dots & 0 \cr
0 &  0 & \dots & 0 & e^{N(\l_{i+}^{(m)}(x)-\l_{i-}^{(m)}(x))} & 0 & \dots & 0 \cr
0 &  0 & \dots & 0 & 0 & 1 & \dots & 0 \cr
\dots &  \dots  & \dots &  \dots  &  \dots  &  \dots  & \dots &  \dots  \cr
0 &  0 & \dots & 0 & 0 & 0 & \dots & 1 \cr
\end{array} \right).
\eeq}

+ for $x \in [z_{2i-1}, r_i^{(1)}]$:
\beq\label{jumpU2}
j_U = \left( \begin{array}{cccccc}
e^{N(\l_{0+}(x)-\l_{0-}(x))} & 1 & 0 & e^{N(\l_{1+}^{(2)}(x)-\l_{0-}(x))}& \dots & e^{N(\l_{l+}^{(b_l)}(x)-\l_{0-}(x))} \cr
0 & e^{N(\l_{1+}^{(0)}(x)-\l_{1-}^{(0)})} & 0 &0& \dots & 0 \cr
0 & 0 & 1 & 0 & \dots & 0 \cr
\dots & \dots & \dots & \dots & \dots & \dots \cr
0 & 0 & 0 &0 & \dots & 1 \cr
\end{array} \right).
\eeq

+ for $x \in [r_i^{(b_i)},z_{2i}]$, $j_U$ is given by:
{\tiny \beq\label{jumpU3}
\left( \begin{array}{cccccccc}
e^{N(\l_{0+}(x)-\l_{0-}(x))} & e^{N(\l_{1+}^{(0)}(x)-\l_{0-}(x))} & \dots & 0& 1 & e^{N(\l_{i+1+}^{(0)}(x)-\l_{0-}(x))}& \dots & e^{N(\l_{l+}^{(b_l)}(x)-\l_{0-}(x))} \cr
0 & 1 & \dots & 0 & 0 & 0 & \dots & 0 \cr
\dots &  \dots  & \dots &  \dots  &  \dots  &  \dots  & \dots &  \dots  \cr
0 &  0 & \dots & 1 & 0 & 0 & \dots & 0 \cr
0 &  0 & \dots & 0 & e^{N(\l_{i+}^{(b_i)}(x)-\l_{i-}^{(b_i)}(x))} & 0 & \dots & 0 \cr
0 &  0 & \dots & 0 & 0 & 1 & \dots & 0 \cr
\dots &  \dots  & \dots &  \dots  &  \dots  &  \dots  & \dots &  \dots  \cr
0 &  0 & \dots & 0 & 0 & 0 & \dots & 1 \cr
\end{array} \right).
\eeq}

+ for $x \in \mathbb{R} \backslash \cup_i [x_{2i-1},x_{2i}]$

\beq
j_U = \left( \begin{array}{ccccc}
1 & e^{N(\l_{1+}^{(0)}(x)-\l_{0-}(x))} & e^{N(\l_{1+}^{(1)}(x)-\l_{0-}(x))} & \dots & e^{N(\l_{l+}^{(b_l)}(x)-\l_{0-}(x))} \cr
0 & 1 & 0 & \dots & 0 \cr
0 & 0 & 1 & \dots & 0 \cr
\dots & \dots & \dots & \dots & \dots \cr
0 & 0 & 0 & \dots & 1 \cr
\end{array} \right).
\eeq

+ for $x \in [x_{2i-1},z_{2i-1}]$, $j_U$ is the same matrix as for $x \in \mathbb{R} \backslash {\displaystyle \bigcup_i} [x_{2i-1},x_{2i}]$
with the term $(i,1)$ of the first line taken equal to 0.

+ for $x \in [z_{2i},x_{2i}]$
$j_U$ is the same matrix as for $x \in \mathbb{R} \backslash {\displaystyle \bigcup_i} [x_{2i-1},x_{2i}]$
with the term $(i,b_i-1)$ of the first line taken equal to 0.

+ for $x \in \Gamma_i^{(m)}$
\bea\label{jumpimaginary}
\left[j_{U}(x)\right]_{(j,h);(j',h')}&=& \delta_{(j,h);(j',h')} \left[1-\delta_{(j,h);(i,m)}
+ \delta_{(j,h);(i,m-1)} \left( e^{N\left(\l_{i+}^{(m-1)}-\l_{i-}^{(m-1)}\right)}-1\right) \right] \cr
& & + \delta_{(j,h);(i,m)}\delta_{(j',h');(i,m-1)} - \delta_{(j,h);(i,m-1)}\delta_{(j',h');(i,m)} \cr
&& - \delta_{(j,h);(i,m-1)} \delta_{(j',h');(i,m-2)} e^{N\left(\l_i^{(m-2)}-\l_{i-}^{(m-1)}\right)}\cr
&& + \delta_{(j,h);(i,m)} \left[ \delta_{(j',h');(i,m+1)} e^{N\left(\l_i^{(m+1)}-\l_{i-}^{(m)}\right)}
- \delta_{(j',h');(i,m-2)} e^{N\left(\l_i^{(m-2)}-\l_{i+}^{(m)}\right)} \right]\cr
\eea
It means that the jump matrix reduces to the identity outside the $2 \times 4$ block:
\beq\label{jumpbloc}
\begin{array}{c|cccc}
 & (i,m-2) & (i,m-1) & (i,m) & (i,m+1) \cr \hline
(i,m-1) & e^{N\left(\l_i^{(m-2)}-\l_{i-}^{(m-1)}\right)} & e^{N\left(\l_{i+}^{(m-1)}-\l_{i-}^{(m-1)}\right)} & -1 & 0 \cr
(i,m) & e^{N\left(\l_i^{(m-2)}-\l_{i+}^{(m)}\right)} & 1 & 0 & e^{N\left(\l_i^{(m+1)}-\l_{i-}^{(m)}\right)} \cr
\end{array}
\eeq

+ for $x \in \Sigma_i$ between $x_{2i-1} + i \mathbb{R}$ and $r_i^{(1)} + i \mathbb{R}$
\beq
\left[j_{U}(x)\right]_{(j,h);(j',h')}:= \delta_{(j,h);(j',h')} + \delta_{(j,h);(i,0)} \delta_{(j',h');(i,1)} e^{N\left(\l_i^{(1)}(x)-\l_i^{(0)}(x)\right)};
\eeq

+ for $x \in \Sigma_i$ between $\Gamma_i^{(m)}$ and $\Gamma_i^{(m+1)}$:
\bea
\left[j_{U}(x)\right]_{(j,h);(j',h')}&:=& \delta_{(j,h);(j',h')} + \delta_{(j,h);(i,m)} \delta_{(j',h');(i,m-1)} e^{N\left(\l_i^{(m-1)}(x)-\l_i^{(m)}(x)\right)}\cr
& & + \delta_{(j,h);(i,m)} \delta_{(j',h');(i,m+1)} e^{N\left(\l_i^{(m+1)}(x)-\l_i^{(m)}(x)\right)};\cr
\eea

+ for $x \in \Sigma_i$ between $r_i^{(b_i)} + i \mathbb{R}$ and $x_{2i} + i \mathbb{R}$
\beq
\left[j_{U}(x)\right]_{(j,h);(j',h')}:= \delta_{(j,h);(j',h')} + \delta_{(j,h);(i,b_i)} \delta_{(j',h');(i,b_i-1)} e^{N\left(\l_i^{(b_i-1)}(x)-\l_i^{(b_i)}(x)\right)}.
\eeq

\end{itemize}

Using the study of the sign of $\Re(\l_i^{(m)})-\Re(\l_j^{(h)})$ performed in the beginning of this section, let us check
that these jump matrices have no exponentially increasing entries when $N \to \infty$. For $x \in \Sigma$
and $x \in \left(\mathbb{R} \backslash {\displaystyle \bigcup_i}[z_{2i-1},z_2i]\right)$, the jump matrices converge to
the identity matrix as $N \to \infty$. For $x \in \Gamma_i^{(m)}$, the jump matrix \ref{jumpimaginary} reduces to the identity
except the block \ref{jumpbloc} which converges to the block
\beq
\begin{array}{c|cccc}
 & (i,m-2) & (i,m-1) & (i,m) & (i,m+1) \cr \hline
(i,m-1) & 0 & 0 & -1 & 0 \cr
(i,m) & 0 & 1 & 0 & 0 \cr
\end{array}
\eeq
exactly as it was encountered in the simpler case studied by \cite{Kuijlaars2}.

Finally, all the non-constant off-diagonal entries of the remaining jump matrices \ref{jumpU1}, \ref{jumpU2} and \ref{jumpU3}
converge to 0 as $N \to \infty$. The non- constant diagonal entries have modulus 1 and are rapidly oscillating as $N \to \infty$.
The third transformation aims to turn them to exponentially decreasing terms.

\subsubsection{Third transformation}
The third transformation aims to cancel the oscillating parts of the jump matrices by opening lenses on $[z_{2i-1},z_{2i}]$
as in the large time case.

Remark hat one can further factorize the matrix $j_U(x)$ for $x \in [r_i^{(m)},r_i^{(m+1)}]$:
\beq
j_U(x) = \widetilde{j}_{US}(x)
j_{S}(x)
\widehat{j}_{US}(x)
\eeq
where, for $x \in [r_i^{(m)},r_i^{(m+1)}]$, one has defined the $k+1\times k+1$ matrices:
\bea
\left[(j_{S})\right]_{(j,h);(j',h')} &:=& (1-\delta_{(j,h);(i,m)}-\delta_{(j,h);(0,0)})\delta_{(j,h);(j',h')} + \cr
&& \qquad \qquad
+ \delta_{(j,h);(0,0)} \delta_{(j',h');(i,m)} - \delta_{(j,h);(i,m)} \delta_{(j',h');(0,0)}+ \cr
&& + \delta_{(j,h);(0,0)} {\displaystyle \sum_{{\tiny \begin{array}{c}(j",h")\neq (0,0),(i,m), \cr (i,m-1),(i,m+1)\cr\end{array}}}}
\delta_{(j',h');(j",h")} e^{N\left(\l_{j'+}^{(h')} - \l_{0-}\right)} -\cr
&& - \delta_{(j,h);(i,m)} {\displaystyle \sum_{{\tiny \begin{array}{c}(j",h")\neq (0,0),(i,m), \cr (i,m-1),(i,m+1)\cr\end{array}}}}
\delta_{(j',h');(j",h")} e^{N\left(\l_{j'+}^{(h')} - \l_{0+}\right)} ,\cr
\eea
\beq
\left[\widetilde{j}_{US}(x)\right]_{(j,h);(j',h')} := \delta_{(j,h);(j',h')} +\delta_{(j,h);(i,m)} \delta_{(j',h');(0,0)} e^{N(\l_0-\l_i^{(m)})_-}
\eeq
and
\beq
\left[\widehat{j}_{US}(x)\right]_{(j,h);(j',h')} :=  \delta_{(j,h);(j',h')} +\delta_{(j,h);(i,m)} \delta_{(j',h');(0,0)} e^{N(\l_0-\l_i^{(m)})_+}  .
\eeq

\begin{figure}
\hspace{3cm}  \includegraphics[width=11cm]{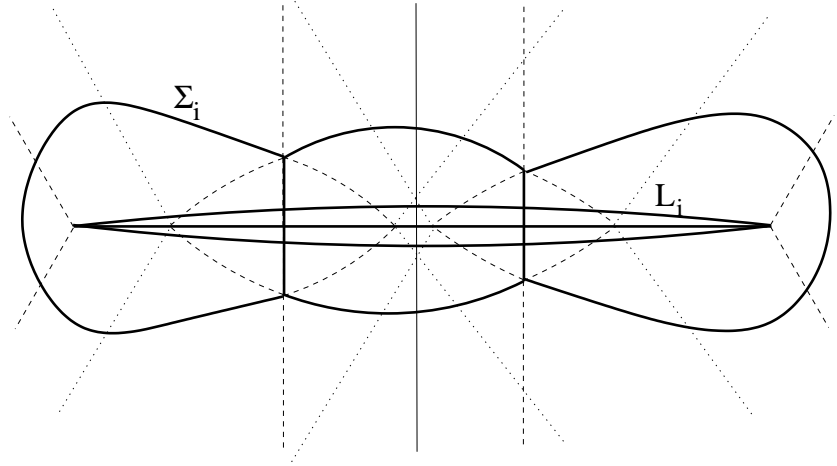}\\
  \caption{Example of lens for the third transformation.}\label{lenseinter}
\end{figure}

As in the large time case, we define a set of $l$ lenses $L_i$ with vertices $z_{2i}$ and $z_{2i-1}$ (see fig.\ref{lenseinter}) which do not
intersect the contour $\Sigma$. Using these lenses, we define the change of variables:
\beq
S(x):= U(x)
\eeq
outside of the lenses and, for $x$ inside the lenses:

\beq
S(x):= \left\{
\begin{array}{l}
U(x) J_{US}^{(up)}(x) \;\; \hbox{in the upper lens region} \cr
U(x) J_{US}^{(down)}(x) \;\; \hbox{in the lower lens region} \cr
\end{array} \right.
\eeq
with
\beq
\left\{\begin{array}{l}
\left[J_{US}^{(up)}\right]_{(j,h);(j',h')} := \delta_{(j,h);(j',h')} - \delta_{(j,h);(i,m)} \delta_{(j',h');(0,0)} e^{N(\l_0-\l_i^{(m)})} \cr
\left[J_{US}^{(down)}\right]_{(j,h);(j',h')} := \delta_{(j,h);(j',h')} + \delta_{(j,h);(i,m)} \delta_{(j',h');(0,0)} e^{N(\l_0-\l_i^{(m)})} \cr
\end{array}\right.
\eeq
for $x$  between $\Gamma_i^{(m)}$ and $\Gamma_i^{(m+1)}$,
\beq
\left\{\begin{array}{l}
\left[J_{US}^{(up)}\right]_{(j,h);(j',h')} := \delta_{(j,h);(j',h')} - \delta_{(j,h);(i,0)} \delta_{(j',h');(0,0)} e^{N(\l_0-\l_i^{(0)})}\cr
\left[J_{US}^{(down)}\right]_{(j,h);(j',h')} := \delta_{(j,h);(j',h')} + \delta_{(j,h);(i,0)} \delta_{(j',h');(0,0)} e^{N(\l_0-\l_i^{(0)})}\cr
\end{array}\right.
\eeq
for $x$ between $z_{2i-1}$ and $\Gamma_i^{(1)}$ and,
\beq
\left\{\begin{array}{l}
\left[J_{US}^{(up)}\right]_{(j,h);(j',h')} := \delta_{(j,h);(j',h')} - \delta_{(j,h);(i,b_i)} \delta_{(j',h');(0,0)} e^{N(\l_0-\l_i^{(b_i)})}\cr
\left[J_{US}^{(down)}\right]_{(j,h);(j',h')} := \delta_{(j,h);(j',h')} + \delta_{(j,h);(i,b_i)} \delta_{(j',h');(0,0)} e^{N(\l_0-\l_i^{(b_i)})}\cr
\end{array}\right.
\eeq
for $x$ between $\Gamma_i^{(b_i)}$ and $z_{2i}$.

We finally get the jump matrices for $S(x)$ out of the discontinuities of $U(x)$ thanks to the factorization of the jump
matrices for $U$ on the real axis.
One has the jumps $S_+(x) = S_-(x) j_S(x)$ with the jump matrix given by:

\begin{itemize}
\item for $x \in ]r_i^{(m)},r_i^{(m+1)}[$, with $i=1,\dots,l$ and $m=0,\dots, b_i$:
\bea
\left[(j_{S})\right]_{(j,h);(j',h')} &:=& (1-\delta_{(j,h);(i,m)}-\delta_{(j,h);(0,0)})\delta_{(j,h);(j',h')}
+ \delta_{(j,h);(0,0)} \delta_{(j',h');(i,m)} - \delta_{(j,h);(i,m)} \delta_{(j',h');(0,0)} \cr
&& + \delta_{(j,h);(0,0)} {\displaystyle \sum_{{\tiny \begin{array}{c}(j",h")\neq (0,0),(i,m), \cr (i,m-1),(i,m+1)\cr\end{array}}}}
\delta_{(j',h');(j",h")} e^{N\left(\l_{j'+}^{(h')} - \l_{0-}\right)}\cr
&& - \delta_{(j,h);(i,m)} {\displaystyle \sum_{{\tiny \begin{array}{c}(j",h")\neq (0,0),(i,m), \cr (i,m-1),(i,m+1)\cr\end{array}}}}
\delta_{(j',h');(j",h")} e^{N\left(\l_{j'+}^{(h')} - \l_{0+}\right)}\cr
\eea

\item for $x \in \mathbb{R} \backslash {\displaystyle \cup_i} [x_{2i-1},x_{2i}]$

\beq
j_S = \left( \begin{array}{ccccc}
1 & e^{N(\l_{1+}^{(0)}(x)-\l_{0-}(x))} & e^{N(\l_{1+}^{(1)}(x)-\l_{0-}(x))} & \dots & e^{N(\l_{l+}^{(b_l)}(x)-\l_{0-}(x))} \cr
0 & 1 & 0 & \dots & 0 \cr
0 & 0 & 1 & \dots & 0 \cr
\dots & \dots & \dots & \dots & \dots \cr
0 & 0 & 0 & \dots & 1 \cr
\end{array} \right).
\eeq

\item for $x \in [x_{2i-1},z_{2i-1}]$, $j_S$ is the same matrix as for $x \in \mathbb{R} \backslash \cup_i [x_{2i-1},x_{2i}]$
with the term $(i,1)$ of the first line taken equal to 0.

\item for $x \in [z_{2i},x_{2i}]$
$j_S$ is the same matrix as for $x \in \mathbb{R} \backslash \cup_i [x_{2i-1},x_{2i}]$
with the term $(i,b_i-1)$ of the first line taken equal to 0.

\item For $x \in \Gamma_i^{(m)}$ between $\tilde{x}_{i,up}^{(m)}$ and $w_i^{(m)}$ and between $\overline{w}_i^{(m)}$ and $\tilde{x}_{i,down}^{(m)}$
\bea\label{jumpimaginary1}
\left[j_{S}(x)\right]_{(j,h);(j',h')}&=& \delta_{(j,h);(j',h')} \left[1-\delta_{(j,h);(i,m)}
+ \delta_{(j,h);(i,m-1)} \left( e^{N\left(\l_{i+}^{(m-1)}-\l_{i-}^{(m-1)}\right)}-1\right) \right] \cr
& & + \delta_{(j,h);(i,m)}\delta_{(j',h');(i,m-1)} - \delta_{(j,h);(i,m-1)}\delta_{(j',h');(i,m)} \cr
&& - \delta_{(j,h);(i,m-1)} \delta_{(j',h');(i,m-2)} e^{N\left(\l_i^{(m-2)}-\l_{i-}^{(m-1)}\right)}\cr
&& + \delta_{(j,h);(i,m)} \left[ \delta_{(j',h');(i,m+1)} e^{N\left(\l_i^{(m+1)}-\l_{i-}^{(m)}\right)}
- \delta_{(j',h');(i,m-2)} e^{N\left(\l_i^{(m-2)}-\l_{i+}^{(m)}\right)} \right]\cr
\eea
It means that the jump matrix reduces to the identity outside the $2 \times 4$ block:
\beq\label{jumpbloc2}
\begin{array}{c|cccc}
 & (i,m-2) & (i,m-1) & (i,m) & (i,m+1) \cr \hline
(i,m-1) & e^{N\left(\l_i^{(m-2)}-\l_{i-}^{(m-1)}\right)} & e^{N\left(\l_{i+}^{(m-1)}-\l_{i-}^{(m-1)}\right)} & -1 & 0 \cr
(i,m) & e^{N\left(\l_i^{(m-2)}-\l_{i+}^{(m)}\right)} & 1 & 0 & e^{N\left(\l_i^{(m+1)}-\l_{i-}^{(m)}\right)} \cr
\end{array}
\eeq

\item For $x \in \Gamma_i^{(m)}$ between $r_i^{(m)}$ and $\tilde{x}_{i,up}^{(m)}$:
\bea\label{jumpimaginary2}
\left[j_{S}(x)\right]_{(j,h);(j',h')}&=& \delta_{(j,h);(j',h')} \left[1-\delta_{(j,h);(i,m)}
+ \delta_{(j,h);(i,m-1)} \left( e^{N\left(\l_{i+}^{(m-1)}-\l_{i-}^{(m-1)}\right)}-1\right) \right] \cr
&& - \delta_{(j,h);(i,m-1)} \delta_{(j',h');(0,0)} e^{N \left(\l_0-\l_{i-}^{(m-1)}\right)} \cr
& & + \delta_{(j,h);(i,m)}\delta_{(j',h');(i,m-1)} - \delta_{(j,h);(i,m-1)}\delta_{(j',h');(i,m)} \cr
&& - \delta_{(j,h);(i,m-1)} \delta_{(j',h');(i,m-2)} e^{N\left(\l_i^{(m-2)}-\l_{i-}^{(m-1)}\right)}\cr
&& + \delta_{(j,h);(i,m)} \left[ \delta_{(j',h');(i,m+1)} e^{N\left(\l_i^{(m+1)}-\l_{i-}^{(m)}\right)}
- \delta_{(j',h');(i,m-2)} e^{N\left(\l_i^{(m-2)}-\l_{i+}^{(m)}\right)} \right]\cr
\eea
i.e. this is the same jump matrix as \ref{jumpimaginary1} except the term $(0,0)$ of the $(i,m-1)$'st line which
is equal to $-e^{N \left(\l_0-\l_{i-}^{(m-1)}\right)}$ instead of 0.

\item For $x \in \Gamma_i^{(m)}$ between $\tilde{x}_{i,down}^{(m)}$ and $r_i^{(m)}$:
\bea
\left[j_{S}(x)\right]_{(j,h);(j',h')}&=& \delta_{(j,h);(j',h')} \left[1-\delta_{(j,h);(i,m)}
+ \delta_{(j,h);(i,m-1)} \left( e^{N\left(\l_{i+}^{(m-1)}-\l_{i-}^{(m-1)}\right)}-1\right) \right] \cr
&& + \delta_{(j,h);(i,m-1)} \delta_{(j',h');(0,0)} e^{N \left(\l_0-\l_{i-}^{(m-1)}\right)} \cr
& & + \delta_{(j,h);(i,m)}\delta_{(j',h');(i,m-1)} - \delta_{(j,h);(i,m-1)}\delta_{(j',h');(i,m)} \cr
&& - \delta_{(j,h);(i,m-1)} \delta_{(j',h');(i,m-2)} e^{N\left(\l_i^{(m-2)}-\l_{i-}^{(m-1)}\right)}\cr
&& + \delta_{(j,h);(i,m)} \left[ \delta_{(j',h');(i,m+1)} e^{N\left(\l_i^{(m+1)}-\l_{i-}^{(m)}\right)}
- \delta_{(j',h');(i,m-2)} e^{N\left(\l_i^{(m-2)}-\l_{i+}^{(m)}\right)} \right]\cr
\eea
i.e. this is the same jump matrix as \ref{jumpimaginary1} except the term $(0,0)$ of the $(i,m-1)$'st line which
is equal to $e^{N \left(\l_0-\l_{i-}^{(m-1)}\right)}$ instead of 0.

\item for $x \in \Sigma_i$ between $x_{2i-1} + i \mathbb{R}$ and $r_i^{(1)} + i \mathbb{R}$
\beq
\left[j_{S}(x)\right]_{(j,h);(j',h')}:= \delta_{(j,h);(j',h')} + \delta_{(j,h);(i,0)} \delta_{(j',h');(i,1)} e^{N\left(\l_i^{(1)}(x)-\l_i^{(0)}(x)\right)};
\eeq

\item for $x \in \Sigma_i$ between $\Gamma_i^{(m)}$ and $\Gamma_i^{(m+1)}$:
\bea
\left[j_{S}(x)\right]_{(j,h);(j',h')}&:=& \delta_{(j,h);(j',h')} + \delta_{(j,h);(i,m)} \delta_{(j',h');(i,m-1)} e^{N\left(\l_i^{(m-1)}(x)-\l_i^{(m)}(x)\right)}\cr
& & + \delta_{(j,h);(i,m)} \delta_{(j',h');(i,m+1)} e^{N\left(\l_i^{(m+1)}(x)-\l_i^{(m)}(x)\right)};\cr
\eea

\item for $x \in \Sigma_i$ between $r_i^{(b_i)} + i \mathbb{R}$ and $x_{2i} + i \mathbb{R}$
\beq
\left[j_{S}(x)\right]_{(j,h);(j',h')}:= \delta_{(j,h);(j',h')} + \delta_{(j,h);(i,b_i)} \delta_{(j',h');(i,b_i-1)} e^{N\left(\l_i^{(b_i-1)}(x)-\l_i^{(b_i)}(x)\right)}.
\eeq

\item for $x\in L_i$  between $\Gamma_i^{(m)}$ and $\Gamma_i^{(m+1)}$
\beq
\left[J_{S}\right]_{(j,h);(j',h')} := \delta_{(j,h);(j',h')} + \delta_{(j,h);(i,m)} \delta_{(j',h');(0,0)} e^{N(\l_0-\l_i^{(m)})}
\eeq

\item for $x \in L_i$ between $z_{2i-1}$ and $\Gamma_i^{(1)}$
\beq
\left[J_{S}\right]_{(j,h);(j',h')} := \delta_{(j,h);(j',h')} + \delta_{(j,h);(i,0)} \delta_{(j',h');(0,0)} e^{N(\l_0-\l_i^{(0)})}
\eeq

\item for $x\in L_i$ between $\Gamma_i^{(b_i)}$ and $z_{2i}$
\beq
\left[J_{S}\right]_{(j,h);(j',h')} := \delta_{(j,h);(j',h')} + \delta_{(j,h);(i,b_i)} \delta_{(j',h');(0,0)} e^{N(\l_0-\l_i^{(b_i)})}
\eeq

\end{itemize}

Now, when $N\to \infty$ all the non-constant terms of the jump matrices tend to 0. We are then left with a model Riemann-Hilbert
problem reduced locally to a $2 \times 2$ matrix problem similar to the one studied in the preceding section. We can thus use
the same parametrix.

\subsubsection{Model Riemann-Hilbert problem}

We proceed exactly as in the preceding section (large time case) and consider the following RH problem (this is the limiting
problem obtained when $N \to \infty$).

\bd
$M$ is solution to the following RH problem:
\begin{itemize}
\item $M: \mathbb{C} \backslash \left(\cup_i [z_{2i-1},z_{2i}] \cup_{i,m} \Gamma_i^{(m)}\right) \to \mathbb{C}^{k+1 \times k+1}$
is analytic

\item $M(x) = I_{k+1 \times k+1} + O\left({1 \over x}\right)$ when $x \to \infty$

\item $M(x)$ satisfies the jumps $M_+(x) = M_-(x) j_M(x)$ with:
\beq
\left[j_M(x)\right]_{(j,h),(j',h')} = \delta_{(j,h),(j',h')}\left(1 - \delta_{(j,h),(0,0)} - \delta_{(j,h),(i,m)}\right)
+ \delta_{(j,h),(0,0)} \delta_{(j',h'),(i,m)} - \delta_{(j',h'),(0,0)} \delta_{(j,h),(i,m)}
\eeq
for $x \in ]r_i^{(m)},r_i^{(m+1)}[$ and
\bea
\left[j_M(x)\right]_{(j,h),(j',h')} &=& \delta_{(j,h),(j',h')}\left(1 - \delta_{(j,h),(i,m-1)} - \delta_{(j,h),(i,m)}\right)
+ \delta_{(j,h),(i,m-1)} \delta_{(j',h'),(i,m)}\cr
&& \quad - \delta_{(j',h'),(i,m-1)} \delta_{(j,h),(i,m)}\cr
\eea
for $x \in \Gamma_i^{(m)}$.

\end{itemize}

\ed

We proceed as in the large time case, following \cite{Kuijlaars2}, to solve this problem: we build a solution under the form of a Lax
matrix associated to the spectral curve. For brevity we will not enter the details of the resolution but only give the result,
the proof is similar to the building of the parametrix away from the branch points in the large time case.

The matrix $M$ takes the following form:
\beq
[M(x)]_{(j,h),(j',h')} = M_j^{(h)}\left(\xi_{j'}^{(h')}(x)\right)
\eeq
where the functions $M_j^{(h)}$ are Baker-Akhiezer functions defined by:
\beq
M_0^{(0)}(x) = {\prod_{(i,m)} \left(x- a_i^{(m)}\right) \over \sqrt{R(x)}}
\eeq
and
\beq
M_j^{(h)}(x) = c_j^{(h)} {\prod_{(i,m)\neq (j,h)} \left(x- a_i^{(m)}\right) \over \sqrt{R(x)}}
\eeq
where
\beq
a_i^{(m)}:= \xi_i^{(m)}(\infty),
\eeq
\beq
R(x):={\displaystyle \prod_{i=1}^l} \left(x-p_{2i-1}\right)\left(x-p_2i\right) {\displaystyle \prod_{i,m}}
\left(x-q_i^{(m)}\right)\left(x-\tilde{q}_i^{(m)}\right)
\eeq
where
\beq
q_i^{(m)}:= \xi_0\left(w_i^{(m)}\right) \qquad \hbox{and} \qquad \tilde{q}_i^{(m)}:= \xi_0\left(\overline{w}_i^{(m)}\right)
\eeq
and
\beq
c_i^{(m)}:=-i \sqrt{n_i^{(m)} \over N} .
\eeq

We now look at the behavior of the solution in the neighborhood of the branch points by approximating it by a local paramerix
build from the Airy function.

Near the real branch points $z_i$, we use the local parametrices $P$ built in the preceding section (see \eq{formPtilde}).
Around the imaginary branch point, we have to build a slightly different parametrix. For this purpose, we can directly
use the parametrix built in section 7 of \cite{}. Let us remind here this result for the parametrix around $w_i^{(m)}$.

Around the branch point $w_i^{(m)}$, let us define a local parametrix $P(x)$ for $x$ inside a disc ${\cal{D}}_i^{(m)}(r)$
of small radius $r$ centered in $w_i^{(m)}$. Studying the behavior of $\l_i^{(m)}(x)-\l_i^{(m-1)}(x)$ inside this disc,
one can define a conformal map $f(x)$ by:
\beq
f(x) := \left[{3 \over 4} \left(\l_i^{(m)}(x)-\l_i^{(m-1)}(x)\right)\right]^{2 \over 3}.
\eeq
One can remark that the neighborhood ${\cal{D}}_i^{(m)}(r)$ is mapped to the complex plane in such a way that:
\beq
\hbox{arg}\left(f(x)\right) = \left\{
\begin{array}{l}
-{2 \pi \over 3} \qquad  \hbox{for} \qquad x \in \Gamma_i^{(m)}\cr
0\,\,\,\,\,\,\,\, \qquad  \hbox{for} \qquad x \in \Sigma_i \; \hbox{in the left half-plane}\cr
{2 \pi \over 3}\,\,\,\,\, \qquad  \hbox{for} \qquad x \in \Sigma_i \; \hbox{in the right half-plane}\cr
\end{array}\right.
\eeq

With these tools, one defines the parametrix $P(x)$ in ${\cal{D}}_i^{(m)}(r)$ by:
\beq\label{paramimagbrach}
P(x) := E(x) \Phi\left( N^{2 \over 3} f(x) \right) D(x)
\eeq
with $D(x)$ is a diagonal matrix defined by
\beq
\left[D(x)\right]_{(j,h);(j',h')}:= \delta_{(j,h);(j',h')} \left[1 + \delta_{(j,h),(i,m-1)} \left( e^{{N \over 2}\left(\l_i^{(m-1)}-\l_i^{(m)}\right)}-1\right)
+ \delta_{(j,h),(i,m)} \left( e^{{N \over 2}\left(\l_i^{(m)}-\l_i^{(m-1)}\right)}-1\right)\right]
\eeq
for $x \in {\cal{D}}_i^{(m)}(r)$, $\Phi(x)$ is the identity matrix except the $2\times 2$ block corresponding to the
intersection of the $(i,m-1)$'th and $(i,m)$'th lines and column which takes the value
\beq
\begin{array}{c|cc}
 & (i,m-1) & (i,m)  \cr \hline
(i,m-1) & -y'_2 & y'_0 \cr
(i,m) & -y_2 & y_0 \cr
\end{array}
\eeq
for $0< \hbox{arg}\left(f(x)\right) < {2 \pi \over 3}$,
\beq
\begin{array}{c|cc}
 & (i,m-1) & (i,m)  \cr \hline
(i,m-1) & -y'_1 & y'_0 \cr
(i,m) & -y_1 & y_0 \cr
\end{array}
\eeq
for $0> \hbox{arg}\left(f(x)\right) > - {2 \pi \over 3}$ and
\beq
\begin{array}{c|cc}
 & (i,m-1) & (i,m)  \cr \hline
(i,m-1) & -y'_2 & y'_1\cr
(i,m) & -y_2 & y_1 \cr
\end{array}
\eeq
for ${2 \pi \over 3}< \hbox{arg}\left(f(x)\right) < {4 \pi \over 3}$ with the functions $y_\alpha$ defined in \eq{defyairy}.
The matrix $E$ is a prefactor used to ensure that the local parametrix can be linked properly to the global parametrix
on the boundary of ${\cal{D}}_i^{(m)}(r)$:
\beq
E(x) := M(x) L(x)^{-1} \qquad \hbox{where} \qquad L(x):= {1 \over 2 \sqrt{\pi}} \widetilde{L}(x) \widehat{L}(x)
\eeq
with $\widetilde{L}$ the diagonal matrix
\beq
\left[\widetilde{L}(x)\right]_{(j,h);(j',h')}:=\delta_{(j,h);(j',h')} \left[1 + \delta_{(j,h);(i,m-1)} \left(N^{1 \over 6} f(x)^{1 \over 4}-1\right)
+ \delta_{(j,h);(i,m)} \left(N^{-{1 \over 6}} f(x)^{-{1 \over 4}}-1\right)\right]
\eeq
and $\widehat{L}$ the identity matrix except for the $2\times 2$ block corresponding to the
intersection of the $(i,m-1)$'th and $(i,m)$'th lines and columns:
\beq
\begin{array}{c|cc}
 & (i,m-1) & (i,m)  \cr \hline
(i,m-1) & i & -1 \cr
(i,m) & i & 1 \cr
\end{array}
\eeq

\subsubsection{Final transformation}

We now have all the ingredients in hand to perform the final transformation and build a solution to our Riemann-Hilbert probem.
Let us define:
\beq
\begin{array}{lr}
R(x) := S(x) M^{-1}(x) & \hbox{outside of the discs surrounding the branch points} \cr
R(x) := S(x) P(x)^{-1} & \hbox{inside of the discs surrounding the branch points} \cr
\end{array}
\eeq
where $P(x)$ is the local parametrix \eq{formPtilde} if one considers a real branch point or \eq{paramimagbrach} if one considers  a complex branch point.

As $N \to \infty$, it is solution to the following RH problem:
\begin{itemize}
\item $R(x)$ is analytic inside the discs surrounding the branch and in $\mathbb{C}\backslash \left( \mathbb{R} \bigcup \Sigma {\displaystyle \bigcup_{i}} L_i {\displaystyle \bigcup_{(i,m)}} \Gamma_i^{(m)}\right)$
outside;

\item its jumps outside the discs are of the form:
\beq
R+ = R- \left({\bf I} + O\left(e^{-\alpha N}\right)\right)
\eeq
with $\alpha >0$ and its jumps on the circles surrounding the branch points are such that:
\beq
R+ = R- \left({\bf I} + O\left({1 \over N}\right)\right).
\eeq

\item as $x \to \infty$:
\beq
R(x) = {\bf I} + O \left({1 \over N}\right).
\eeq

\end{itemize}

As in the large time case this implies the behavior of $R$:
\beq
R(x) = {\bf I} + O \left({1 \over N(|x|+1)}\right)
\eeq
as $N \to \infty$ (see \cite{Kuijlaars2}).

\subsection{Asymptotics of the kernel and universality}

Using this final transformation as well as the properties of the parametrices, one easily shows theorems
\ref{thmeandensity1}, \ref{thkernelbulk} and \ref{thkerneledge} by
mimicking the proof of the large time case.


\section{Perspectives}
\label{secconclu}

We used the Riemann-Hilbert approach to study the asymptotic behavior of $N$ non-intersecting Brownian bridges and show the
universality of their correlation functions as $N \to \infty$. Nevertheless, we did not study the critical configurations corresponding
to a spectral curve where two real branch points merge. One expects to find a universal behavior given by the
Pearcy kernel as it was found in the particular case studied by \cite{Kuijlaars3}. This critical case may be addressed by the
same techniques and is left to a forthcoming work.
However, as the length of these notes witnesses, the RH approach is long and requires a global information on the spectral curve
to finally give local results. It would be interesting to obtain these results in a more direct way using, for example, a
direct steepest-analysis as performed by Adler and Van Moerbeke \cite{AVM}.

\vs

{\large \bf Acknowledgements}

\vs

The author would like to thank A.B.J. Kuijlaars for a critical reading of a first draft of this paper. This research is supported
by the Enigma European network MRT-CT-2004-5652 through a post-doctoral fellowship and benefited also of the support of
the ANR project G\'{e}om\'{e}trie et int\'{e}grabilit\'{e} en physique math\'{e}matique ANR-05-BLAN-0029-01, the Enrage European
network MRTN-CT-2004-005616, the European Science Foundation through the Misgam program and the French and Japanese governments
through PAI Sakura.


\begin{thebibliography}{99}

\bibitem{AVM} M. Adler, P. Van Moerbeke, ``Looking for a non-symmetric Pearcy process'', in preparation.

\bibitem{Bleher} P.M. Bleher, ``Lectures on Random Matrix Models: The Riemann-Hilbert Approach'', arXiv:0801.1858.

\bibitem{Kuijlaars1} P.M. Bleher and A.B.J. Kuijlaars, ``Large n limit of gaussian random matrices with external source, Part I'',
{\em Commun. Math. Phys.} {\bf 252}(2004),43.

\bibitem{Kuijlaars2} A.I. Aptekarev, P.M. Bleher and A.B.J. Kuijlaars, ``Large n limit of gaussian random matrices with external source, Part II'',
{\em Commun. Math. Phys.} {\bf 259}(2005),367.

\bibitem{Kuijlaars3} P.M. Bleher and A.B.J. Kuijlaars, ``Large n limit of gaussian random matrices with external source, Part III:
Double scaling limit'',
{\em Commun. Math. Phys.} {\bf 270}(2006),481.


\bibitem{KD1} E. Daems and A.B.J. Kuijlaars, ``A Christoffel-Darboux formula for multiple orthogonal polynomials'',
{\em J. Approx. Theory} {\bf 130}(2004),190.



\bibitem{EOinvariants} B.Eynard and N. Orantin, ``Invariants of algebraic curves and topological expansion'',
{\em Communication in Number Theory and Physics} {\bf vol.1 n°2}, math-ph/0702045.

\bibitem{KD2} V. Lysov, F. Wielonsky, ``Strong asymptotics for
multiple Laguerre polynomials'',
{\em Constructive Approximation} {\bf 28}(2008), 61.

\bibitem{Pastur} L.A. Pastur, ``The spectrum of random matrices'' (Russian), {\em Teoret. Mat. Fiz.} {\bf 10} (1972), 102.

\bibitem{TW} C. Tracy and H. Widom, ``Nonintersecting Brownian excursions'',
math/0607321.


\bibitem{PZ1} P. Zinn-Justin, ``Random Hermitian matrices in an external field'', {\em Nucl. Phys.} {\bf B 497} (1997), 725.

\bibitem{PZ2} P. Zinn-Justin, ``Universality of orrelation functions of Hermitian random matrices in an external field'',
{\em Comm. Math. Phys.} {\bf 194} (1998), 631.

\end{thebibliography}
\end{document}